# Tunable magnetism and electron correlation in Titanium-based Kagome metals RETi$_3$Bi$_4$ (RE = Yb, Pr, and Nd) by rare-earth engineering

Long Chen[1, 2, #], Ying Zhou[1, 2, #], He Zhang[1, 2, #], Xuecong Ji[1, 2, #], Ke Liao[1, 2], Yu Ji[1, 2], Ying Li[1], Zhongnan Guo[3], Xi Shen[1], Richeng Yu[1, 2, 4], Xiaohui Yu[1, 2, 4, *], Hongming Weng[1, 2, 4, *], Gang Wang[1, 2, 4, *]

[1] Beijing National Laboratory for Condensed Matter Physics, Institute of Physics, Chinese Academy of Sciences, Beijing 100190, China

[2] University of Chinese Academy of Sciences, Beijing 100049, China

[3] Department of Chemistry, School of Chemistry and Biological Engineering, University of Science and Technology Beijing, Beijing 100083, China

[4] Songshan Lake Materials Laboratory, Dongguan, Guangdong 523808, China

[#] These authors contributed equally to this work.

[*] Corresponding author. Email: yuxh@iphy.ac.cn; hmweng@iphy.ac.cn; gangwang@iphy.ac.cn.

**Abstract:**

Rare-earth engineering is an effective way to introduce and tune the magnetism in topological Kagome magnets, which has been acting as a fertile platform to investigate the quantum interactions between geometry, topology, spin, and correlation. Here we report the structure and properties of three newly discovered Titanium-based Kagome metals RETi$_3$Bi$_4$ (RE = Yb, Pr, and Nd) with various magnetic states. They crystallize in the orthogonal space group *Fmmm* (No.69), where slightly distorted Ti Kagome lattice, RE triangular lattice, Bi honeycomb and triangular lattices stack along the *a* axis. By changing the rare earth atoms on RE zag-zig chains, the magnetism can be tuned from nonmagnetic YbTi$_3$Bi$_4$ to short-range ordered PrTi$_3$Bi$_4$ ($T_{anomaly}$ ~ 8.2 K), and finally to ferromagnetic NdTi$_3$Bi$_4$ ($T_c$ ~ 8.5 K). The measurements of resistivity and specific heat capacity demonstrate an evolution of electron correlation and density of states near the Fermi level with different rare earth atoms. *In-situ* resistance measurements of NdTi$_3$Bi$_4$ under high pressure further reveal a potential relationship between the electron correlation and ferromagnetic ordering temperature. These results highlight RETi$_3$Bi$_4$ as another family of topological Kagome magnets to explore nontrivial band topology and exotic phases in Kagome materials.

# 1 Introduction

Topological Kagome magnets and superconductors have attracted tremendous attention in the past decades, which have been serving as a fundamental platform to investigate the quantum interactions between geometry, topology, spin and correlation [1]. From Heisenberg's view [2], magnetic ions on the Kagome lattice with antiferromagnetic interaction would exhibit geometrical frustration [3], resulting in possible quantum-spin-liquid states [4], fractionalized excitations [5] and the absence of ordinary magnetic order [6, 7]. When focusing on the electronic structure of Kagome lattices, the strong electron correlation originated from the flat band would stabilize a ferromagnetic (FM) ground state [8]. Besides, with Dirac fermions and van Hove singularities in its electronic structure, Kagome lattices would also show nontrivial topological properties [9, 10] and multiple long-range orders with instabilities of Fermi surface [11]. By introducing the magnetic degrees of freedom, correlated topological band structures can be realized in Kagome lattice, known as topological Kagome magnets. Theoretically, the inclusion of spin-orbit coupling in the Kagome lattice would open gaps at the Dirac point and the touching point connecting flat-band and quadratic band, making the Kagome lattice a quantum spin Hall insulator with nontrivial $Z_2$ topological invariants [12, 13]. Thus, Kagome magnets with out-of-plane FM ordering would be a Chern insulator with chiral edge states, which usually exhibits exotic quantum properties such as quantized anomalous Hall effect [14] and giant anomalous Hall conductance [15, 16]. Moreover, large in-plane magnetization would close the gaps in topological Kagome magnets, acting as a tuning parameter for topological phases [1]. Considering the strong or weak interlayer interaction, Weyl point [17-19] or three-dimensional quantum anomalous Hall effect [20] is anticipated to show up in topological Kagome magnets. Due to the exotic quantum properties, topological Kagome magnets have great potential applications for next-generation electronic technology, and it is essential to explore new topological Kagome magnets and their properties.

Basically, there are two strategies for the introduction of magnetism in Kagome materials, putting magnetic ions on the Kagome lattice or intercalating magnetic ions between stacked Kagome interlayers. Plenty of topological Kagome magnets have been reported with magnetic transition metal ions on the Kagome lattice, such as $Mn_3X$ (X = Sn, Ge) with Mn-Kagome lattice [21-23], $Co_3Sn_2S_2$ [24-26] and CoSn [27, 28] with Co-Kagome lattice, $Fe_3Sn_2$ [16, 29, 30] and FeSn [31-33] with Fe-Kagome lattice, most of which have strong in-plane magnetization. In $REMn_6Sn_6$ (RE = Gd-Tm, Lu) with Mn-Kagome lattice, various magnetic states can realized by substitution of different rare-earth atoms, named rare earth engineering, which results quite different quantum transport behaviors [34, 35]. In particular, $TbMn_6Sn_6$ shows an out-of-plane magnetization and stands out as a Chern insulator with zero-field anomalous Hall, anomalous Nernst and anomalous thermal Hall effects [10, 35-37]. By replacing the magnetic Mn-Kagome lattice with nonmagnetic V-Kagome lattice, weak magnetic couplings among the local $4f$ moments of various rare earth atoms in $REV_6Sn_6$ not only result in various magnetic orderings [38-40], but also topologically nontrivial Dirac surface states ($GdV_6Sn_6$ and $HoV_6Sn_6$ [41-43]), quantum oscillation ($YV_6Sn_6$ [42]), quantum critical behavior ($YbV_6Sn_6$ [44]), and charge density waves ($ScV_6Sn_6$ [45-47]). Similar to Kagome magnets, rare earth engineering has also been proved to largely influence the magnetism, magnetoresistance or topological properties of square-net compounds [48-50]. The close relationship between rare-earth magnetism and topological electronic structure highlights that rare-earth engineering is an effective way to tune the exotic topological phases [35]. Recently, the discovery of $AM_3X_5$ (A = K, Rb, Cs; M = Ti, V; X = Sb, Bi) family with V/Ti-Kagome lattice [51, 52] have attracted a lot of attention due to the combination of charge density wave states, superconductivity, orbital-selective nematic order or electronic nematicity, and nontrivial band topology [53-60]. Upon applying hydrostatic pressure or carrier doping, multiple SC domes can be induced in $AV_3Sb_5$, showing the competition between CDW and superconductivity [61-67]. By introducing RE atoms in VSb or TiBi interlayers, Vanadium-based Kagome magnets begin to emerge [68], showing a FM-like ground state in $EuV_3Sb_4$. However, the Titanium-based Kagome magnets are still rare, i.e. only $LaTi_3Bi_4$, $CeTi_3Bi_4$ and $EuTi_3Bi_4$ [69] are reported but without systematic investigations.

Based on rare earth engineering, we report the synthesis and physical properties of three newly discovered Titanium-based Kagome metals $RETi_3Bi_4$ (RE = Yb, Pr, and Nd). $RETi_3Bi_4$ (RE = Yb, Pr, and Nd) all crystallize in the orthogonal space group *Fmmm* with stacking layers composed of Ti-Kagome lattice, RE triangular lattice, Bi honeycomb and triangular lattices. Compared with $AM_3X_5$, the Ti-Kagome lattice in $RETi_3Bi_4$ is slightly distorted and zig-zag RE chains are formed in REBi bilayers, making it possible to show various magnetic states with various RE atoms. The combined measurements of magnetism, resistivity and heat capacity show an evolution from the nonmagnetic $YbTi_3Bi_4$ to short-range ordered $PrTi_3Bi_4$ with an anomaly around 8.2 K, and finally to FM $NdTi_3Bi_4$ with $T_c$ = 8.5 K. The power-law fitting of resistivity at low temperature suggests an enhancement of electron correlation by changing RE atoms from Yb to Pr and Nd, which is consistent with the increasing of density of states near the Fermi level according to the Debye model fitting of heat capacity. By applying pressure, the FM order is slightly suppressed with an increasing value of power in $NdTi_3Bi_4$, showing potential relationship between the magnetism and electron correlation. Therefore, $RETi_3Bi_4$ (RE = Yb, Pr, and Nd) should be another family of topological Kagome magnets to investigate the interplay between topologically nontrivial feature, magnetism and electron correlation.

**2 Experimental**

**Single Crystal Growth.** $RETi_3Bi_4$ (RE = Yb, Pr and Nd) single crystals were grown by a high-temperature solution method using Bi as flux. The as-received Yb/Pr/Nd ingot (Alfa, 99.9%) was cut into small pieces, then mixed with Ti powder (99.95%, Alfa Aesar) and Bi granules (99.999%, Sinopharm) with a molar ratio of Yb/Pr/Nd : Ti : Bi = 2 : 4 : 12 in a fritted alumina crucible set (Canfield crucible set) [70] and sealed in a fused-silica ampoule at vacuum. The ampoule was heated to 1073 K over 15 h, held at the temperature for 24 h, and then slowly cooled down to 873 K at a rate of 2 K/h. At 873 K, hexagonal-shaped, shiny-silver single crystals with size up to 5 mm × 5 mm × 0.5 mm were separated from the remaining liquid by centrifuging the ampoule. Considering the possible air sensitivity of the surface, all manipulations and specimen preparation for structure characterization and property measurements were handled in an argon-filled glove box.

**Structure Characterization and Composition Analysis.** X-ray diffraction data were obtained using a PANalytical X'Pert PRO diffractometer (Cu $K_α$ radiation, λ = 1.54178 Å) operated at 40 kV voltage and 40 mA current with a graphite monochromator in a reflection mode (2θ = 5°–100°, step size = 0.017°). Indexing and Rietveld refinement were performed using the DICVOL91 and FULLPROF programs [71]. Single crystal X-ray diffraction (SCXRD) data were collected using a Bruker D8 VENTURE with Mo $K_α$ radiation (λ = 0.71073Å) at 280 K for $PrTi_3Bi_4$ and $NdTi_3Bi_4$. The structure was solved using a direct method and refined with the Olex2 [72] and Jana2020 [73] package. Due to the air-sensitivity and two-dimensional feature of $YbTi_3Bi_4$, the crystal structure cannot be easily determined by SCXRD. The focused ion beam (FIB) method was used to prepare thin specimens of YbTi$_3$Bi$_4$ with a thickness of ~50 nm for scanning transmission electron microscopy (STEM) analysis. The crystal structure of YbTi$_3$Bi$_4$ was characterized by high-angle annular dark-field (HAADF) images obtained using a JEOL ARM-200F transmission electron microscope with double Cs correctors (CEOS) for the condenser lens and objective lens. The available spatial resolution is better than 78 picometers at 200 kV. The morphology and analyses of elements were characterized using a scanning electron microscope (SEM-4800, Hitachi) equipped with an electron microprobe analyzer for semiquantitative elemental analysis in energy-dispersive spectroscopy (EDS) mode. Three spots in three different locations were measured on each crystal using EDS.

**Physical Property Measurements.** Temperature-dependent magnetic susceptibility were measured using a vibrating sample magnetometer system (VSM, Quantum Design), under a magnetic field (0.5 T) parallel ($H // bc$) and normal ($H \perp bc$) to the *bc* plane using both the zero-field-cooling (ZFC) and field-cooling (FC) protocols. Only magnetic susceptibility in FC protocol was measured for a larger field (1 T), and the field-dependent magnetization curves were measured under the magnetic field up to 7 T parallel and perpendicular to the *bc* plane. The resistivity and heat capacity measurements were carried out using a physical property measurement system (Quantum Design, 9 T). The resistivity was measured using the standard four-probe configuration with the applied current (about 2 mA) parallel to the *bc* plane. Heat capacity measurement was carried out at temperature ranging from 2.2 K to 200 K at high vacuum (0.01 μbar). To protect the sample from air and moisture, thin film of N-type grease (~0.05 mg) was spread to cover the sample in an argon-filled glove box first and then the sample was mounted on the square plate of specialized heat capacity pucks in air.

***In-situ* high pressure resistance measurement.** The standard four-probe resistance measurement under high pressure was carried out to obtain the high-pressure electrical transport properties of the NdTi$_3$Bi$_4$ sample at high-pressure synergetic measurement station, synergetic extreme condition user facility. A Cu-Be alloy diamond anvil cell (DAC) with 300 *μ*m diamond culet in diameter was used. For the sample chamber, a 100 *μ*m diameter hole was drilled in a rhenium gasket with a c-BN insulating layer, which was then filled with KBr as the pressure transmitting medium. Long-stick shaped single crystal sample NdTi$_3$Bi$_4$ was loaded into the sample chamber in an Argon atmosphere glovebox. Four Pt electrodes were contacted to the sample in four-probe configuration. The pressure of the sample chamber could be calibrated by ruby fluorescence method at room temperature.

## 3 Results and Discussion

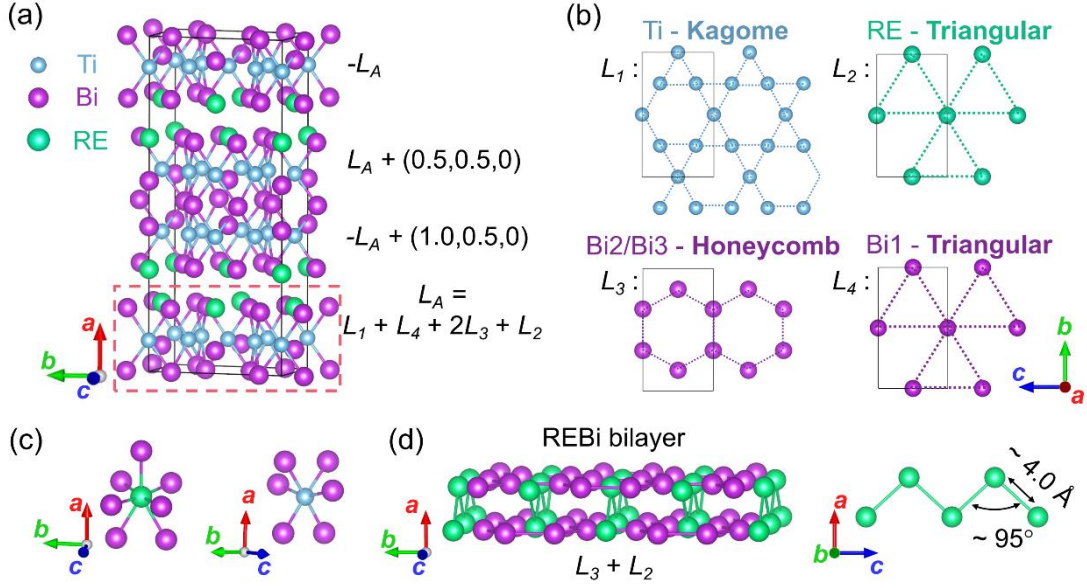

**Figure 1. Crystal structure of RETi$_3$Bi$_4$ (RE = Yb, Pr and Nd).** (a) Crystal structure of RETi$_3$Bi$_4$ (RE = Yb, Pr and Nd) and correspond layer construction operations. The dashed rectangular denotes the construction layer $L_A$. (b) The components of $L_A$ viewed along the $a$ axis, including Ti Kagome-like lattice ($L_1$), RE triangular lattice ($L_2$), Bi honeycomb-like lattices ($L_3$) and Bi triangular lattice ($L_4$). (c) The coordination environment of RE and Ti, respectively. (d) The REBi bilayer and zig-zag chain of RE atoms.

Unlike the hexagonal prototypes AM$_3$X$_5$ (A = K, Rb, Cs; M = Ti, V; X = Sb, Bi) [51, 52, 60], all RETi$_3$Bi$_4$ crystallize in the orthogonal space group *Fmmm* (No.69) with $a$ = 24.9(4) Å, $b$ = 10.3(4) Å, $c$ = 5.9(2) Å for YbTi$_3$Bi$_4$; $a$ = 24.9668(114) Å, $b$ = 10.3248(49) Å, $c$ = 5.9125(26) Å for PrTi$_3$Bi$_4$; $a$ = 24.9523(87) Å, $b$ = 10.3327(27) Å, $c$ = 5.9009(18) Å for NdTi$_3$Bi$_4$, and $\alpha = \beta = \gamma = 90°$ (Table S1-S5). As shown in **Figure 1**a, the crystal structure is composed of the alternating stacking of an initial building layer ($L_A$) along the $a$ axis. Each unit cell contains four building layers, which connected each other by a mirror symmetry along the $a$ axis, a (0.5, 0.5, 0) translation symmetry and their joint symmetry. The initial building layer ($L_A$) can be further resolved into a Ti Kagome-like lattice, a RE triangular lattice, two Bi honeycomb-like lattices and a Bi triangular lattice (Figure 1b). There are two distinct differences between the crystal structure of AM$_3$X$_5$ and RETi$_3$Bi$_4$, i.e. (i) the increase of the $a$ axis owing to the stacking of (ii) distorted TiBi layer with distorted Ti Kagome lattice. Compared with ATi$_3$Bi$_5$, both the Ti Kagome and Bi triangular lattice of RETi$_3$Bi$_4$ show inequivalent bonds and are not strictly in the $bc$ plane (Figure 1c), resulting in both in-plane and out-of-plane distortion of TiBi layer. Separately speaking, both the Bi honeycomb lattice and Ti Kagome lattice would show interesting topological properties, and the triangular lattice of magnetic ions would unavoidably give rise to the geometrical frustration. As each building layer are of great interesting in exploring their topological properties and magnetism, RETi$_3$Bi$_4$ is anticipated to be a good platform exhibit unique properties. Similar to LnV$_3$Sb$_4$ (Ln = Yb, Eu) [68], the bilayer of REBi plane forms a quasi-one-dimensional zig-zag chain of RE atoms (Figure 1d), on which the magnetic states could be tunable by rare-earth engineering.

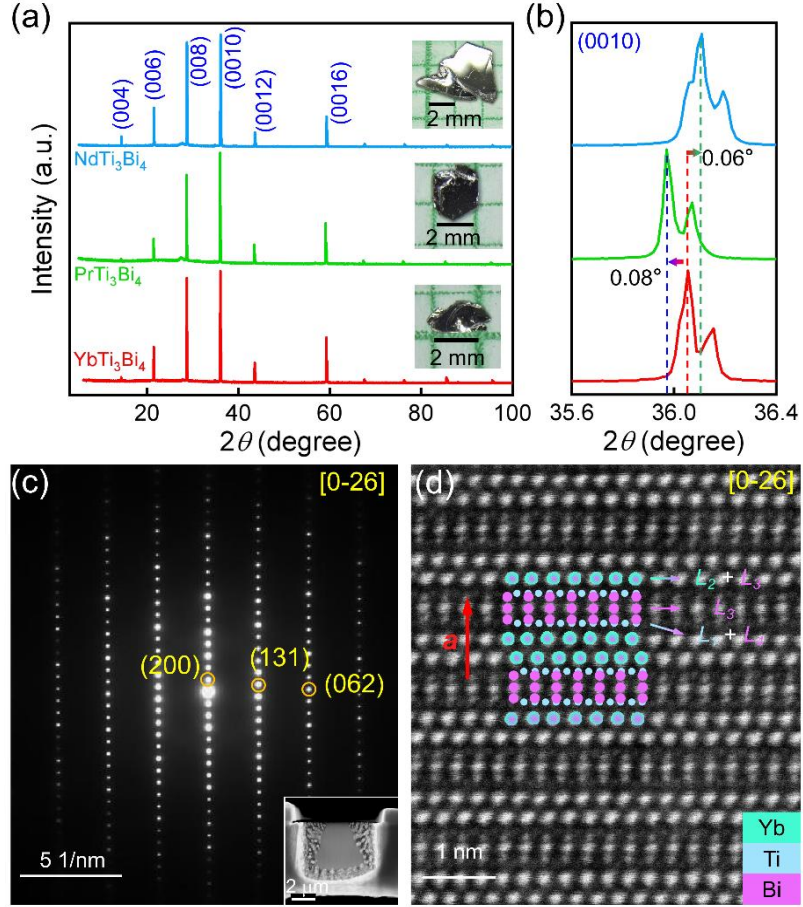

**Figure 2. Single Crystals of RETi$_3$Bi$_4$ (RE = Yb, Pr and Nd).** (a) X-ray diffraction pattern of as-grown RETi$_3$Bi$_4$ (RE = Yb, Pr and Nd) single crystals, showing (00*l*) (*l* = even integers) reflections. The insets are the corresponding optical photographs of RETi$_3$Bi$_4$ (RE = Yb, Pr and Nd) single crystals. (b) The enlarged (0010) peak of RETi$_3$Bi$_4$ (RE = Yb, Pr and Nd) single crystals. (c) The SAED pattern of YbTi$_3$Bi$_4$ along the [0-26] zone-axis. The inset is the thin specimen for YbTi$_3$Bi$_4$ etched by FIB. (d) The HAADF image collected along the [0-26] zone-axis, where the cyan, blue, and purple balls represent Yb, Ti, and Bi atoms, respectively

The crystal structure is further confirmed by the XRD pattern, which show a strong preferential orientation of (00*l*) (*l* = even integer) reflections (**Figure 2**a). Based on the position (~14.3°) of (004) diffraction peaks, the distance between corresponding structural units along the *a* axis is determined to be about 25 Å, close to the results of SCXRD. As shown in the insets of Figure 2a, the as-grown single crystals of RETi$_3$Bi$_4$ are all plate-like flakes with shiny metal luster, indicating a clear quasi-two-dimensional feature. In particular, the single crystals of PrTi$_3$Bi$_4$ and NdTi$_3$Bi$_4$ show clear distorted hexagonal shape, corresponding to the existence of Kagome-like lattice. In all synthesized crystals, the single crystals of PrTi$_3$Bi$_4$ and NdTi$_3$Bi$_4$ look thicker than that of YbTi$_3$Bi$_4$, showing possible stronger out-of-plane interaction than that of YbTi$_3$Bi$_4$. Figure 2b shows the enlarged (0010) peaks for RETi$_3$Bi$_4$, where the peak position of PrTi$_3$Bi$_4$ shift 0.08° to lower angle and NdTi$_3$Bi$_4$ shift 0.06° to higher angle than that of YbTi$_3$Bi$_4$, indicating a shortest distance between the structural units for NdTi$_3$Bi$_4$ and the longest for PrTi$_3$Bi$_4$. The different (0010) peak positions corresponds well with the lattice parameter derived from SCXRD, where $a$(NdTi$_3$Bi$_4$) < $a$(YbTi$_3$Bi$_4$) < $a$(PrTi$_3$Bi$_4$). Considering the cationic radius relation $r$(Yb$^{3+}$) = 0.0858 nm < $r$(Yb$^{2+}$) = 0.093 nm <

$r(Nd^{3+}) = 0.0995$ nm $< r(Pr^{3+}) = 0.1013$ nm, the abnormal enhancement of *a* axis in YbTi$_3$Bi$_4$ might be attributed the weakness of out-of-plane interaction between YbBi bilayer with Yb$^{2+}$ ions rather than Yb$^{3+}$ ions. As shown in the selected area electron diffraction (SAED) pattern and HAADF image of YbTi$_3$Bi$_4$ collected along the [0-26] zone-axis (Figure 2c, d), the construction layers can be clearly resolved along the *a* axis. In particular, the Ti atoms and Bi atoms are not on the same straight line, further confirming the out-of-plane distortion of TiBi layers. The chemical composition is determined to be RE : Ti : Bi ~ 1 : 3 : 4 with a homogenous distribution according to the result of EDS (Figure S1-S2 and Table S6-S8).

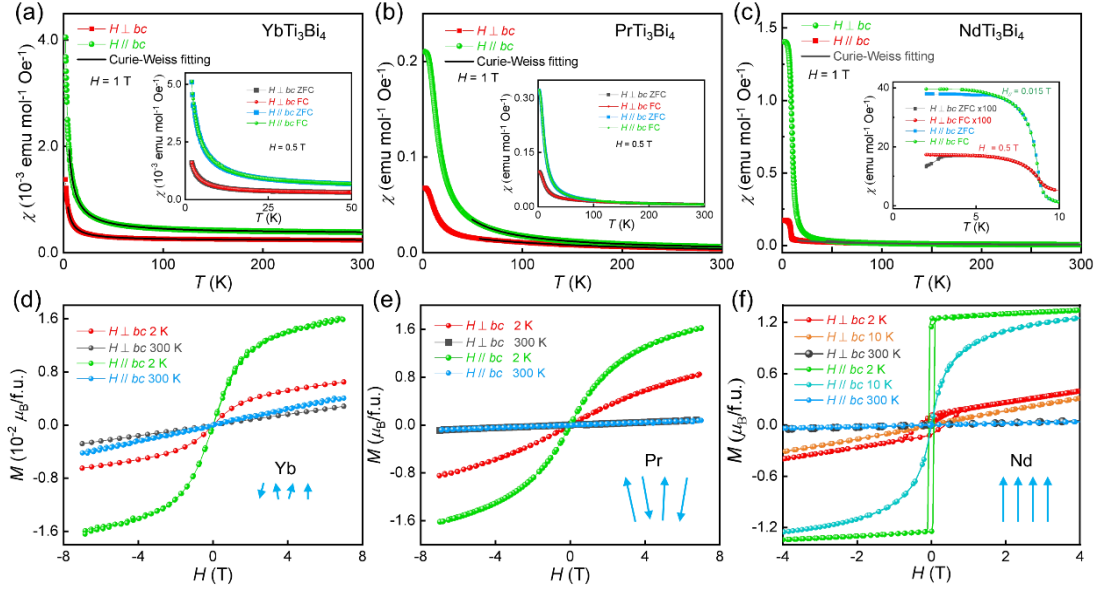

**Figure 3. Various magnetic states of RETi$_3$Bi$_4$ (RE = Yb, Pr and Nd).** Temperature-dependent magnetic susceptibility of (a) YbTi$_3$Bi$_4$, (b) PrTi$_3$Bi$_4$, and (c) NdTi$_3$Bi$_4$ with large magnetic field (1 T) parallel and perpendicular to the *bc* plane. The black lines are Curie-Weiss fitting curves. The insets show correspond ZFC and FC curves at small magnetic field. Field-dependent magnetization curves for (d) YbTi$_3$Bi$_4$, (e) PrTi$_3$Bi$_4$, and (f) NdTi$_3$Bi$_4$ with magnetic fields parallel and perpendicular to the *bc* plane. The insets show schematic alignment of spin moments in the magnetic ground state of RETi$_3$Bi$_4$.

**Figure 3**a-c shows the magnetic susceptibility of RETi$_3$Bi$_4$ under field parallel (*H // bc*) and normal (*H* ⊥ *bc*) to the *bc* plane. The in-plane (*H // bc*) magnetic susceptibilities are larger than the out-of-plane (*H* ⊥ *bc*) magnetic susceptibilities, showing a quasi-2D feature of all RETi$_3$Bi$_4$. For YbTi$_3$Bi$_4$, the ZFC and FC curves under a small magnetic field (0.5 T) show no bifurcation above 2 K, indicating nonmagnetic transitions. Though without bifurcation of the ZFC and FC curves under small magnetic field (0.5 T) above 2 K, a saturation tendency exists at low temperature in the magnetic susceptibility of PrTi$_3$Bi$_4$ for both *H // bc* and *H* ⊥ *bc*, possibly indicating a short-range order around $T_{anomaly}$ ~ 8.2 K (Figure S3). By contrast, a clear bifurcation can be resolved for NdTi$_3$Bi$_4$, which corresponds to a possible FM ordering with transition temperature $T_c$ = 8.5 K.

All the magnetic susceptibility decrease quickly around 20 K and follow the Curie-Weiss law at high temperature (black lines), $\chi = \chi_0 + C/(T - \theta)$, where $\chi_0$ is the temperature-independent contribution including the diamagnetic contribution of the orbital magnetic moment, *C* the Curie constant, and $\theta$ the Curie temperature. The effective moment can be further calculated from the equation $\mu_{eff} = \sqrt{\frac{8C}{n}}$, where $n$ is the number of magnetic atom. For YbTi$_3$Bi$_4$, almost zero Curie temperature (~0.5 K) and small effective moment (< 0.3 $\mu_B$/Yb) were found for both $H$ // $bc$ and $H \perp bc$, indicating a possible non-magnetic ground state with zero-spin Yb$^{2+}$ (0 $\mu_B$) state. For PrTi$_3$Bi$_4$, a positive but close-to-zero Curie temperature (~2.5 K) was observed and the effective moment is calculated to be 3.67(3) $\mu_B$/Pr for $H$ // $bc$, close to that of the theoretical moment of Pr$^{3+}$ (3.58 $\mu_B$) spin state. Such result may suggest weak magnetic exchange couplings between different Pr triangular bilayers along the $a$ axis. Changing the magnetic field into $H \perp bc$, the large negative Curie temperature (-100 K) suggest strong antiferromagnetic interactions between moments in the $bc$ plane, and the overestimation of the effective moment (4.72 $\mu_B$/Pr) should be attributed to the geometrical spin frustration of Pr triangular lattice. By substitution Pr with Nd, a small negative Curie temperature (-3.6 K) for $H$ // $bc$ would also suggest weak magnetic exchange couplings between different Nd triangular bilayers along the $a$ axis. But the reduction of Curie temperature (-36.7 K) for $H \perp bc$ may corresponds to a weakening of antiferromagnetic interactions in the $bc$ plane, which might result in the emergence of FM order in NdTi$_3$Bi$_4$. The derived effective moments (~ 3.6 $\mu_B$) of NdTi$_3$Bi$_4$ are close to the theoretical moment of Nd$^{3+}$ (3.62 $\mu_B$) spin moment. Details of the Curie-Weiss fittings can be seen in Table S9 and Figure S4.

The magnetization curves of YbTi$_3$Bi$_4$ and PrTi$_3$Bi$_4$ show positive slopes and then approach saturation around 7 T without any hysteresis (Figure 3d, e), suggesting no FM order or spin-glass state at least above 2 K. The saturation moment of YbTi$_3$Bi$_4$ is 0.016 $\mu_B$ for $H$ // $bc$ and 0.006 $\mu_B$ for $H \perp bc$, further confirming the non-magnetic feature ($4f^{14}$, $\mu_{sat} = 0$ $\mu_B$). The saturation moment of PrTi$_3$Bi$_4$ is 1.620 $\mu_B$ for $H$ // $bc$ and 0.855 $\mu_B$ $H \perp bc$, of which the former one is close to the saturation moment of Pr$^{3+}$ ($4f^2$, $\mu_{sat} = 2.0$ $\mu_B$). The large difference between the in-plane and out-of-plane saturation moment should be attributed to the large anisotropy originated from the quasi-2D structure. For NdTi$_3$Bi$_4$, obvious hysteresis loops exist at 2 K and gradually vanish beyond $T_c$ (Figure 3f and Figure S5), further confirming the existence of FM order. The smaller in-plane coercive magnetic field ($H_{c//}$ ~ 500 Oe << $H_{c\perp}$ ~ 5000 Oe) and larger saturation moment ($\mu_{sat-//}$ ~ 1.426 $\mu_B$ > $\mu_{sat-\perp}$ ~ 0.587 $\mu_B$) suggest that the easy plane is the $bc$ plane. Thus, various magnetic ground states have been realized in titanium-based Kagome magnets RETi$_3$Bi$_4$ by rare earth engineering, which so far contains nonmagnetic YbTi$_3$Bi$_4$, possibly short-range ordered PrTi$_3$Bi$_4$, and FM NdTi$_3$Bi$_4$.

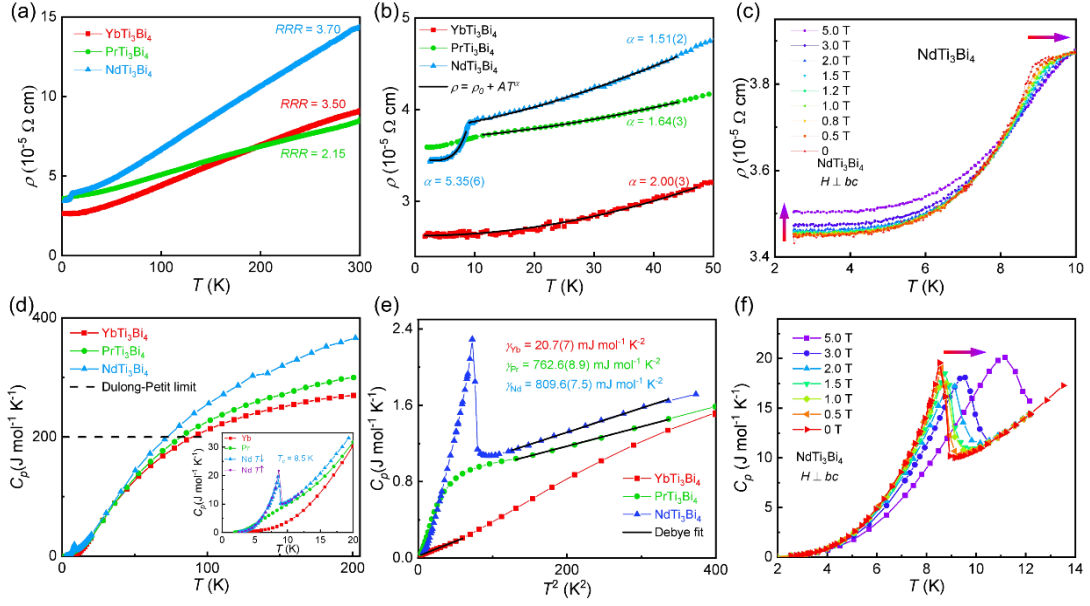

**Figure 4. Electrical transport and specific heat capacity of RETi$_3$Bi$_4$ (RE = Yb, Pr and Nd).** (a) Temperature-dependent in-plane resistivity ($\rho_{bc}$) of RETi$_3$Bi$_4$ (RE = Yb, Pr and Nd). (b) Power-law fittings of in-plane resistivity at low temperature. (c) In-plane resistivity of NdTi$_3$Bi$_4$ at low temperature under different magnetic field. (d) Temperature-dependent specific heat of RETi$_3$Bi$_4$ (RE = Yb, Pr and Nd). The inset shows zoomed specific heat at low temperature. The black dashed line shows the Dulong-Petit limit of RETi$_3$Bi$_4$ compounds. (e) $C_P$ vs $T^2$ plot for RETi$_3$Bi$_4$ (RE = Yb, Pr and Nd) and corresponds fittings (black lines) using Debye model. (f) Zoomed specific heat of NdTi$_3$Bi$_4$ at low temperature under different magnetic field.

**Figure 4**a shows the temperature-dependent in-plane resistivity of RETi$_3$Bi$_4$, which monotonically decreases with decreasing of temperature, showing a metallic-like behavior (See Figure S6 for the metallic electronic structure of PrTi$_3$Bi$_4$ and NdTi$_3$Bi$_4$). The residual-resistance ratios ($RRR=\rho_{300\,K}/\rho_{10\,K}$) are calculated to be 3.50 for YbTi$_3$Bi$_4$, 2.15 for PrTi$_3$Bi$_4$ and 3.70 for NdTi$_3$Bi$_4$, hinting the good crystallinity for RETi$_3$Bi$_4$ single crystals. With current (2 mA) being applied in the *bc* plane, the resistivity of YbTi$_3$Bi$_4$ shows no obvious anomaly down to 2 K. Unlike YbTi$_3$Bi$_4$, the in-plane resistivity of PrTi$_3$Bi$_4$ at low temperature shows an anomaly enhancement with a broaden peak (5.2 K ~ 9.7 K) in its first derivative curve (Figure S7), which might be correlated with the possible short-range order. Due to the existence of FM order in NdTi$_3$Bi$_4$, the resistivity shows a distinct drop around $T_c$ = 8.5 K and then approach a saturation at 2 K. Similiar to AV$_3$Sb$_5$ or ATi$_3$Bi$_5$, the resistivity for RETi$_3$Bi$_4$ below 50 K can be well fitted using power-law: $\rho = \rho_0 + AT^\alpha$, where the value of power $\alpha$ is dependent on the dominant scattering mechanism (Table S10). Usually, $\alpha$ takes the value of 3/2 for diffusive electron motion caused by strong electron correlation [74], 2 for moderate electron-electron scattering [75, 76], known as the Fermi liquid behavior [77], 3 for dominant *s−d* scattering or electron-magnon scattering [78-80], and 5 for electron-phonon coupling [81]. As shown in Figure 4b, the value of power ($\alpha$) is fitted to be 2.00(3) for YbTi$_3$Bi$_4$, 1.64(3) for PrTi$_3$Bi$_4$ and 1.51(2) for NdTi$_3$Bi$_4$, indicating a strengthen of electron correlation by changing rare earth atoms from Yb to Pr, and then to Nd. In particular, the resistivity for NdTi$_3$Bi$_4$ below $T_c$ can be also fitted using power-law with $\alpha$ = 5.35(6), showing strong electron-phono coupling below $T_c$. By applying magnetic field perpendicular to the *bc* plane, the resistivity

below $T_c$ shows a slight enhancement up to 5 T with almost unchanged transition temperature (Figure 4c), which is a typical characteristic of FM order.

Figure 4d shows the temperature-dependent specific heat of RETi$_3$Bi$_4$, where no anomaly in the entire measured temperature range, a broaden anomaly around 8.2 K, and a sharp peak at $T_c$ = 8.5 K was observed for YbTi$_3$Bi$_4$, PrTi$_3$Bi$_4$, and NdTi$_3$Bi$_4$, respectively. All the specific heat capacity of RETi$_3$Bi$_4$ single crystals slightly exceeds the Dulong–Petit limit (3NR ~ 200 J mol$^{-1}$ K$^{-1}$, black dashed lines) at higher temperature, which is attributed to the more prominent phonon contribution at higher temperature of N-type grease used for protecting the samples [82-84]. For NdTi$_3$Bi$_4$, the peak at 8.5 K is reversible by cooling ($T\downarrow$) and warming ($T\uparrow$), indicating no structural transition. The low-temperature (2 K - 5 K for YbTi$_3$Bi$_4$, 12 K - 18 K for PrTi$_3$Bi$_4$ and NdTi$_3$Bi$_4$) specific heat capacity of RETi$_3$Bi$_4$ is fitted by the Debye model $C = \gamma T + \beta T^3$, where $\gamma T$ and $\beta T^3$ represent the contribution from the electron and lattice, respectively (Figure 4e). The Sommerfeld coefficients $\gamma$ proportional to the density of states (DOSs) around Fermi level ($E_F$) [85] ($\gamma \propto g(E_F)$) are determined to be 20.7(7) mJ K$^{-2}$ per formula for YbTi$_3$Bi$_4$, 762.6(8.9) mJ K$^{-2}$ per formula for PrTi$_3$Bi$_4$, and 809.6(7.5) mJ K$^{-2}$ per formula for NdTi$_3$Bi$_4$. For the insulating character of N-type grease, $\gamma$ correlated with the electron contribution will not be affected at low temperature. The quite large $\gamma$ indicate large DOSs near $E_F$ for both PrTi$_3$Bi$_4$ and NdTi$_3$Bi$_4$, making it possible for them to show strong electron correlation, which is consistent with the results of resistivity. The λ-shaped peak of specific heat capacity for NdTi$_3$Bi$_4$ broadens and shifts toward higher temperature upon increasing the out-of-plane magnetic field (Figure 4f), further confirming the FM behavior of NdTi$_3$Bi$_4$. Taking the nonmagnetic YbTi$_3$Bi$_4$ as the reference of specific heat capacity contributed by electron and phonon, the contribution of magnon and magnetic entropy of PrTi$_3$Bi$_4$ and NdTi$_3$Bi$_4$ are calculated (Figure S8), suggesting possible Pr$^{3+}$ and Nd$^{3+}$ states in PrTi$_3$Bi$_4$ and NdTi$_3$Bi$_4$, respectively.

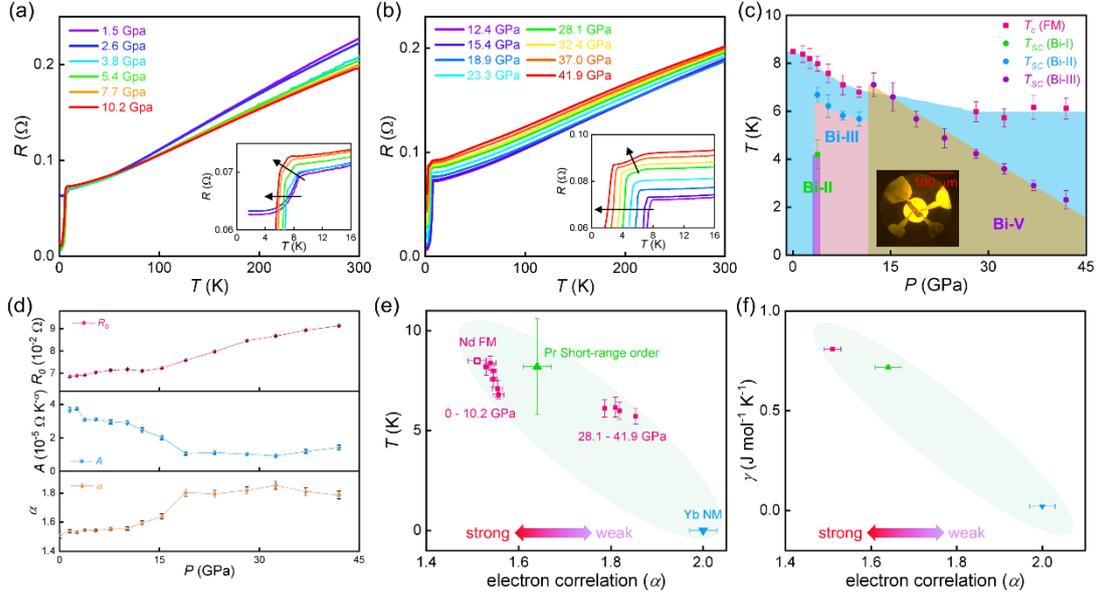

**Figure 5. Magnetism and electron correlation tuned by pressure.** Temperature-dependent resistance of NdTi$_3$Bi$_4$ under high pressure from (a) 1.5 GPa to 10.2 GPa, (b) 12.4 GPa to 41.9 GPa. The insets show the resistance at low temperature from 1.7 K to 12 K. The arrows show the suppression trend of FM ordering temperature and superconducting transition temperature. (c)

Pressure dependence of FM ordering temperature and superconducting transition temperature. The inset shows the optical photograph four probe configuration in the diamond-anvil cell. (d) The fitted parameters of resistance at low temperature using power-law fitting $R = R_0 + AT^\alpha$ under different pressure. (e) The evolution of FM ordering temperature with the value of power ($\alpha$), which represents the strength of electron correlation. The ordering temperature for nonmagnetic $YbTi_3Bi_4$ is taken to be zero and $PrTi_3Bi_4$ taken to be 8.2(2.0) K for comparison. (f) The evolution of Sommerfeld coefficient ($\gamma$) with the value of power ($\alpha$).

*In situ* resistance measurements of $NdTi_3Bi_4$ in the pressure range of 1.5 GPa - 41.9 GPa were performed by using a diamond-anvil cell (**Figure 5**a-b). Up to 10.2 GPa, the small drop related to FM order can be observed in resistance with a superconducting-like transition happened beyond 3.8 GPa (inset of Figure 5a), where both the FM ordering temperature and superconducting transition temperature are suppressed as pressure increases (Figure S9). By increasing the pressure beyond 10.2 GPa, the drop related to the FM order cannot be resolved because the superconducting transition temperature suddenly steps higher (Figure 5c). Until the pressure reach 28.1 GPa, the drop related to the FM order shows up again owing to the quicker suppression of superconducting transition. However, zero-resistance state does not exist in all the measured curves, which may hint an extrinsically superconducting transition. By carefully evaluating the upper critical field (Figure S10), the superconducting signals should be attributed to three pressure-induced Bi phases (Bi-II, Bi-III, and Bi-V) [86], which might come from the remaining Bi-flux droplets or be synthesized under pressure in this Bi-rich compounds. Though with the superconducting side phase, the resistance beyond FM ordering temperature (10 K - 45 K) still follow the power-law, $R = R_0 + AT^\alpha$. Figure 5d and Table S11 show the pressure dependence of fitted parameters, where residue resistance ($R_0$) increases and correlation coefficient ($A$) decreases with the increasing pressure. In particular, the value of power ($\alpha$) increase from 1.51(2) to 1.82(3) when pressure increases from 0 GPa to 32.4 GPa, showing a decreasing trend of electron correlation. Figure 5e-f summarize the evolution of FM ordering temperature and Sommerfeld coefficient with the value of power, where the stronger electron correlation corresponds to a higher FM ordering temperature and higher DOSs near $E_F$. This result may hint us to find Kagome magnets with higher FM ordering temperature in a system with much stronger electron correlation and much higher DOSs near $E_F$.

**4 Conclusion**

In summary, three newly discovered Titanium-based Kagome metals $RETi_3Bi_4$ (RE = Yb, Pr, and Nd) are reported. By rare earth engineering, various magnetic states have been realized, including nonmagnetic $YbTi_3Bi_4$, paramagnetic-like $PrTi_3Bi_4$ with an anomaly around $T_{anomaly} \sim 8.2$ K, and FM $NdTi_3Bi_4$ with $T_c = 8.5$ K. Both the electron correlation and density of states around Fermi level are hinted to be tuned by different rare earth atoms, indicating rare-earth engineering as an effective way to explore new materials and tune exotic phases. This work also extends the $RETi_3Bi_4$ family with diverse magnetic states by rare earth engineering, providing a new platform to investigate the interplay between topologically nontrivial feature, magnetism and electron correlation. Further calculations and experimental characterizations, like detailed investigation of anisotropic magnetism or magnetoresistance, angle-resolved photoemission spectroscopy, and scanning tunneling microscopy or spectroscopy, are all desired to unravel the possible exotic topological phases of Titanium-based Kagome metals $RETi_3Bi_4$.

**Supporting information**

Table S1-S5 show the crystal structures of RETi$_3$Bi$_4$ (RE = Yb, Pr and Nd) single crystals. Table S6-S8 show the elemental composition of RETi$_3$Bi$_4$ (RE = Yb, Pr and Nd) single crystals. Table S9 shows the fitted parameters of magnetic susceptibility for RETi$_3$Bi$_4$ (RE = Yb, Pr and Nd) using Curie-Weiss law. Table S10 shows the fitted parameters of resistivity using power law and that of specific heat capacity using Debye model for RETi$_3$Bi$_4$ (RE = Yb, Pr and Nd). Table S11 shows the fitted parameters of resistance for NdTi$_3$Bi$_4$ under high pressure using power law.

Figure S1-S2 show the SEM images and chemical composition of RETi$_3$Bi$_4$ (RE = Yb, Pr and Nd) single crystals. Figure S3 shows magnetic susceptibility of PrTi$_3$Bi$_4$ at low temperature. Figure S4 shows magnetic susceptibility and magnetization of NdTi$_3$Bi$_4$ at low temperature. Figure S5 shows the $1/\chi$ *vs* $T$ plot of RETi$_3$Bi$_4$ (RE = Yb, Pr and Nd) single crystals. Figure S6 shows the metallic electronic structure of PrTi$_3$Bi$_4$ and NdTi$_3$Bi$_4$. Figure S7 shows resistivity of PrTi$_3$Bi$_4$ at low temperature. Figure S8 shows the specific heat capacity contributed by magnon of PrTi$_3$Bi$_4$ and NdTi$_3$Bi$_4$. Figure S9 shows the first derivative curves of resistance (d$R$/d$T$) for NdTi$_3$Bi$_4$ under high pressure. Figure S10 shows pressure-induced superconducting transition related to different Bi phases under high pressure.

**Author information**

**Notes**

The authors declare no competing financial interest.

**Acknowledgements**

L. Chen, Y. Zhou, H. Zhang and X.C. Ji contributed equally to this work. L. Chen, Y. Zhou and G. Wang would like to thank Prof. X. L. Chen, Prof. J. P. Hu, Prof. T. Qian at Institute of Physics, Chinese Academy of Sciences, and Prof. J. Ma at Shanghai Jiao Tong University for helpful discussions. This work was partially supported by the National Key Research and Development Program of China (Grant Nos. 2018YFE0202600 and 2017YFA0302902), the National Natural Science Foundation of China (Grant No. 51832010 and 11888101), and the Key Research Program of Frontier Sciences, Chinese Academy of Sciences (Grant No. QYZDJ-SSW-SLH013).

**References**

[1] J.X. Yin, B. Lian, M.Z. Hasan, Topological kagome magnets and superconductors, Nature, 612 (2022) 647-657.

[2] W. Heisenberg, Zur Theorie des Ferromagnetismus, Zeitschrift für Physik, 49 (1928) 619-636.

[3] I. Syôzi, Statistics of Kagomé Lattice, Progress of Theoretical Physics, 6 (1951) 306-308.

[4] S. Yan, D.A. Huse, S.R. White, Spin-liquid ground state of the S = 1/2 Kagome Heisenberg antiferromagnet, Science, 332 (2011) 1173-1176.

[5] T.H. Han, J.S. Helton, S. Chu, D.G. Nocera, J.A. Rodriguez-Rivera, C. Broholm, Y.S. Lee, Fractionalized excitations in the spin-liquid state of a Kagome-lattice antiferromagnet, Nature, 492 (2012) 406-410.

[6] P.W. Anderson, Resonating valence bonds: A new kind of insulator?, Materials Research Bulletin, 8 (1973) 153-160.

[7] C. Broholm, R.J. Cava, S.A. Kivelson, D.G. Nocera, M.R. Norman, T. Senthil, Quantum spin liquids, Science, 367 (2020) eaay0668.

[8] A. Mielke, Ferromagnetic ground states for the Hubbard model on line graphs, Journal of Physics A: Mathematical and General, 24 (1991) L73.

[9] L. Ye, M. Kang, J. Liu, F. von Cube, C.R. Wicker, T. Suzuki, C. Jozwiak, A. Bostwick, E. Rotenberg, D.C. Bell, L. Fu, R. Comin, J.G. Checkelsky, Massive Dirac fermions in a ferromagnetic kagome metal, Nature, 555 (2018) 638-642.

[10] J.X. Yin, W. Ma, T.A. Cochran, X. Xu, S.S. Zhang, H.-J. Tien, N. Shumiya, G. Cheng, K. Jiang, B. Lian, Z. Song, G. Chang, I. Belopolski, D. Multer, M. Litskevich, Z.J. Cheng, X.P. Yang, B. Swidler, H. Zhou, H. Lin, T. Neupert, Z. Wang, N. Yao, T.-R. Chang, S. Jia, M. Zahid Hasan, Quantum-limit Chern topological magnetism in $TbMn_6Sn_6$, Nature, 583 (2020) 533-536.

[11] W.S. Wang, Z.Z. Li, Y.Y. Xiang, Q.-H. Wang, Competing electronic orders on Kagome lattices at van Hove filling, Phys. Rev. B, 87 (2013) 115135.

[12] H.M. Guo, M. Franz, Topological insulator on the kagome lattice, Phys. Rev. B, 80 (2009) 113102.

[13] E. Tang, J.W. Mei, X.G. Wen, High-Temperature Fractional Quantum Hall States, Physical Review Letters, 106 (2011) 236802.

[14] G. Xu, B. Lian, S.C. Zhang, Intrinsic Quantum Anomalous Hall Effect in the Kagome Lattice $Cs_2LiMn_3F_{12}$, Physical Review Letters, 115 (2015) 186802.

[15] E. Liu, Y. Sun, N. Kumar, L. Muechler, A. Sun, L. Jiao, S.-Y. Yang, D. Liu, A. Liang, Q. Xu, J. Kroder, V. Süß, H. Borrmann, C. Shekhar, Z. Wang, C. Xi, W. Wang, W. Schnelle, S. Wirth, Y. Chen, S.T.B. Goennenwein, C. Felser, Giant anomalous Hall effect in a ferromagnetic kagome-lattice semimetal, Nature Physics, 14 (2018) 1125-1131.

[16] T. Kida, L.A. Fenner, A.A. Dee, I. Terasaki, M. Hagiwara, A.S. Wills, The giant anomalous Hall effect in the ferromagnet $Fe_3Sn_2$—a frustrated kagome metal, Journal of Physics: Condensed Matter, 23 (2011) 112205.

[17] H. Weyl, Elektron und Gravitation. I, Zeitschrift für Physik, 56 (1929) 330-352.

[18] X. Wan, A.M. Turner, A. Vishwanath, S.Y. Savrasov, Topological semimetal and Fermi-arc surface states in the electronic structure of pyrochlore iridates, Phys. Rev. B, 83 (2011) 205101.

[19] A.A. Burkov, L. Balents, Weyl Semimetal in a Topological Insulator Multilayer, Physical Review Letters, 107 (2011) 127205.

[20] B.I. Halperin, Possible States for a Three-Dimensional Electron Gas in a Strong Magnetic Field, Japanese Journal of Applied Physics, 26 (1987) 1913.

[21] S. Tomiyoshi, Y. Yamaguchi, Magnetic Structure and Weak Ferromagnetism of $Mn_3Sn$ Studied by Polarized Neutron Diffraction, Journal of the Physical Society of Japan, 51 (1982) 2478-2486.

[22] H. Yang, Y. Sun, Y. Zhang, W.-J. Shi, S.S.P. Parkin, B. Yan, Topological Weyl semimetals in the chiral antiferromagnetic materials $Mn_3Ge$ and $Mn_3Sn$, New J. Phys., 19 (2017) 015008.

[23] A.K. Nayak, J.E. Fischer, Y. Sun, B. Yan, J. Karel, A.C. Komarek, C. Shekhar, N. Kumar, W. Schnelle, J. Kübler, C. Felser, S.S.P. Parkin, Large anomalous Hall effect driven by a nonvanishing Berry curvature in the noncolinear antiferromagnet $Mn_3Ge$, Sci. Adv., 2 (2016) e1501870.

[24] Q. Wang, Y. Xu, R. Lou, Z. Liu, M. Li, Y. Huang, D. Shen, H. Weng, S. Wang, H. Lei, Large intrinsic anomalous Hall effect in half-metallic ferromagnet $Co_3Sn_2S_2$ with magnetic Weyl fermions, Nature Communications, 9 (2018) 3681.

[25] N. Morali, R. Batabyal, P.K. Nag, E. Liu, Q. Xu, Y. Sun, B. Yan, C. Felser, N. Avraham, H. Beidenkopf, Fermi-arc diversity on surface terminations of the magnetic Weyl semimetal $Co_3Sn_2S_2$, Science, 365 (2019) 1286-1291.

[26] D.F. Liu, E.K. Liu, Q.N. Xu, J.L. Shen, Y.W. Li, D. Pei, A.J. Liang, P. Dudin, T.K. Kim, C. Cacho, Y.F. Xu, Y. Sun, L.X. Yang, Z.K. Liu, C. Felser, S.S.P. Parkin, Y.L. Chen, Direct observation of the spin–orbit coupling effect in magnetic Weyl semimetal $Co_3Sn_2S_2$, npj Quantum Mater., 7 (2022) 11.

[27] Z. Liu, M. Li, Q. Wang, G. Wang, C. Wen, K. Jiang, X. Lu, S. Yan, Y. Huang, D. Shen, J.-X. Yin, Z. Wang, Z. Yin, H. Lei, S. Wang, Orbital-selective Dirac fermions and extremely flat bands in frustrated kagome-lattice metal CoSn, Nature Communications, 11 (2020) 4002.

[28] M. Kang, S. Fang, L. Ye, H.C. Po, J. Denlinger, C. Jozwiak, A. Bostwick, E. Rotenberg, E. Kaxiras, J.G. Checkelsky, R. Comin, Topological flat bands in frustrated kagome lattice CoSn, Nature Communications, 11 (2020) 4004.

[29] L.A. Fenner, A.A. Dee, A.S. Wills, Non-collinearity and spin frustration in the itinerant kagome ferromagnet $Fe_3Sn_2$, Journal of Physics: Condensed Matter, 21 (2009) 452202.

[30] S. Fang, L. Ye, M.P. Ghimire, M. Kang, J. Liu, M. Han, L. Fu, M. Richter, J. van den Brink, E. Kaxiras, R. Comin, J.G. Checkelsky, Ferromagnetic helical nodal line and Kane-Mele spin-orbit coupling in kagome metal $Fe_3Sn_2$, Phys. Rev. B, 105 (2022) 035107.

[31] B.C. Sales, J. Yan, W.R. Meier, A.D. Christianson, S. Okamoto, M.A. McGuire, Electronic, magnetic, and thermodynamic properties of the kagome layer compound FeSn, Phys. Rev. Materials, 3 (2019) 114203.

[32] M. Kang, L. Ye, S. Fang, J.-S. You, A. Levitan, M. Han, J.I. Facio, C. Jozwiak, A. Bostwick, E. Rotenberg, M.K. Chan, R.D. McDonald, D. Graf, K. Kaznatcheev, E. Vescovo, D.C. Bell, E. Kaxiras, J. van den Brink, M. Richter, M. Prasad Ghimire, J.G. Checkelsky, R. Comin, Dirac fermions and flat bands in the ideal kagome metal FeSn, Nature Materials, 19 (2020) 163-169.

[33] Z. Lin, C. Wang, P. Wang, S. Yi, L. Li, Q. Zhang, Y. Wang, Z. Wang, H. Huang, Y. Sun, Y. Huang, D. Shen, D. Feng, Z. Sun, J.-H. Cho, C. Zeng, Z. Zhang, Dirac fermions in antiferromagnetic FeSn kagome lattices with combined space inversion and time-reversal symmetry, Phys. Rev. B, 102 (2020) 155103.

[34] G. Venturini, B.C.E. Idrissi, B. Malaman, Magnetic properties of $RMn_6Sn_6$ (R = Sc, Y, Gd−Tm, Lu) compounds with $HfFe_6Ge_6$ type structure, Journal of Magnetism and Magnetic Materials, 94 (1991) 35-42.

[35] W. Ma, X. Xu, J.X. Yin, H. Yang, H. Zhou, Z.J. Cheng, Y. Huang, Z. Qu, F. Wang, M.Z. Hasan, S. Jia, Rare Earth Engineering in $RMn_6Sn_6$ (R=Gd-Tm, Lu) Topological Kagome Magnets, Physical Review Letters, 126 (2021) 246602.

[36] X. Xu, J.X. Yin, W. Ma, H.J. Tien, X.-B. Qiang, P.V.S. Reddy, H. Zhou, J. Shen, H.Z. Lu, T.R. Chang, Z. Qu, S. Jia, Topological charge-entropy scaling in kagome Chern magnet $TbMn_6Sn_6$, Nature Communications, 13 (2022) 1197.

[37] H. Zhang, J. Koo, C. Xu, M. Sretenovic, B. Yan, X. Ke, Exchange-biased topological transverse thermoelectric effects in a Kagome ferrimagnet, Nature Communications, 13 (2022) 1091.

[38] J. Lee, E. Mun, Anisotropic magnetic property of single crystals $RV_6Sn_6$ (R=Y, Gd-Tm, Lu), Phys. Rev. Materials, 6 (2022) 083401.

[39] E. Rosenberg, J.M. DeStefano, Y. Guo, J.S. Oh, M. Hashimoto, D. Lu, R.J. Birgeneau, Y. Lee, L. Ke, M. Yi, J.-H. Chu, Uniaxial ferromagnetism in the kagome metal $TbV_6Sn_6$, Phys. Rev. B, 106 (2022) 115139.

[40] X. Zhang, Z. Liu, Q. Cui, Q. Guo, N. Wang, L. Shi, H. Zhang, W. Wang, X. Dong, J. Sun, Z. Dun, J. Cheng, Electronic and magnetic properties of intermetallic kagome magnets $RV_6Sn_6$ (R=Tb-Tm), Phys. Rev. Materials, 6 (2022) 105001.

[41] S. Peng, Y. Han, G. Pokharel, J. Shen, Z. Li, M. Hashimoto, D. Lu, B.R. Ortiz, Y. Luo, H. Li, M. Guo, B. Wang, S. Cui, Z. Sun, Z. Qiao, S.D. Wilson, J. He, Realizing Kagome Band Structure in Two-Dimensional Kagome Surface States of $RV_6Sn_6$ (R=Gd, Ho), Physical Review Letters, 127 (2021) 266401.

[42] G. Pokharel, S.M.L. Teicher, B.R. Ortiz, P.M. Sarte, G. Wu, S. Peng, J. He, R. Seshadri, S.D. Wilson, Electronic properties of the topological kagome metals $YV_6Sn_6$ and $GdV_6Sn_6$, Phys. Rev. B, 104 (2021) 235139.

[43] Y. Hu, X. Wu, Y. Yang, S. Gao, N.C. Plumb, A.P. Schnyder, W. Xie, J. Ma, M. Shi, Tunable topological Dirac surface states and van Hove singularities in kagome metal $GdV_6Sn_6$, Sci. Adv., 8 (2022) eadd2024.

[44] K. Guo, J. Ye, S. Guan, S. Jia, Triangular Kondo lattice in $YbV_6Sn_6$ and its quantum critical behavior in a magnetic field, Phys. Rev. B, 107 (2023) 205151.

[45] T. Hu, H. Pi, S. Xu, L. Yue, Q. Wu, Q. Liu, S. Zhang, R. Li, X. Zhou, J. Yuan, D. Wu, T. Dong, H. Weng, N. Wang, Optical spectroscopy and band structure calculations of the structural phase transition in the vanadium-based kagome metal $ScV_6Sn_6$, Phys. Rev. B, 107 (2023) 165119.

[46] H.W.S. Arachchige, W.R. Meier, M. Marshall, T. Matsuoka, R. Xue, M.A. McGuire, R.P. Hermann, H. Cao, D. Mandrus, Charge Density Wave in Kagome Lattice Intermetallic $ScV_6Sn_6$, Physical Review Letters, 129 (2022) 216402.

[47] X. Zhang, J. Hou, W. Xia, Z. Xu, P. Yang, A. Wang, Z. Liu, J. Shen, H. Zhang, X. Dong, Y. Uwatoko, J. Sun, B. Wang, Y. Guo, J. Cheng, Destabilization of the Charge Density Wave and the Absence of Superconductivity in $ScV_6Sn_6$ under High Pressures up to 11 GPa, Materials, 15 (2022) 7372.

[48] H. Chen, J. Gao, L. Chen, G. Wang, H. Li, Y. Wang, J. Liu, J. Wang, D. Geng, Q. Zhang, J. Sheng, F. Ye, T. Qian, L. Chen, H. Weng, J. Ma, X. Chen, Topological crystalline insulator candidate ErAsS with hourglass Fermion and magnetic-tuned topological phase transition, Adv. Mater. (Weinheim, Ger.), DOI 10.1002/adma.202110664(2022) 2110664.

[49] B. Kang, Z. Liu, D. Zhao, L. Zheng, Z. Sun, J. Li, Z. Wang, T. Wu, X. Chen, Giant negative magnetoresistance beyond Chiral anomaly in topological material $YCuAs_2$, Adv. Mater. (Weinheim, Ger.), 34 (2022) 2201597.

[50] L. Chen, Y. Gu, Y. Wang, Y. Zhou, K. Liao, Y. Pan, X. Wu, Y. Li, Z. Wang, Y. Ma, Z. Guo, J. Ma, D. Su, J. Hu, G. Wang, Large negative magnetoresistance beyond chiral anomaly in topological insulator candidate $CeCuAs_2$ with spin-glass-like behavior, The Innovation Materials, DOI 10.59717/j.xinn-mater.2023.100011(2023) 100011-100011-100011-100018.

[51] B.R. Ortiz, L.C. Gomes, J.R. Morey, M. Winiarski, M. Bordelon, J.S. Mangum, I.W.H. Oswald, J.A. Rodriguez-Rivera, J.R. Neilson, S.D. Wilson, E. Ertekin, T.M. McQueen, E.S. Toberer, New Kagome prototype materials: discovery of $KV_3Sb_5$, $RbV_3Sb_5$, and $CsV_3Sb_5$, Phys. Rev. Materials, 3 (2019) 094407.

[52] D. Werhahn, B.R. Ortiz, A.K. Hay, S.D. Wilson, R. Seshadri, D. Johrendt, The kagomé metals $RbTi_3Bi_5$ and $CsTi_3Bi_5$, Z. Naturforsch. B, 77 (2022) 757-764.

[53] S. Cho, H. Ma, W. Xia, Y. Yang, Z. Liu, Z. Huang, Z. Jiang, X. Lu, J. Liu, Z. Liu, J. Li, J. Wang, Y. Liu, J. Jia, Y. Guo, J. Liu, D. Shen, Emergence of new van Hove singularities in the charge density wave state of a topological Kagome metal $RbV_3Sb_5$, Physical Review Letters, 127 (2021) 236401.

[54] B.R. Ortiz, P.M. Sarte, E.M. Kenney, M.J. Graf, S.M.L. Teicher, R. Seshadri, S.D. Wilson, Superconductivity in the $Z_2$ kagome metal $KV_3Sb_5$, Phys. Rev. Materials, 5 (2021) 034801.

[55] Q.W. Yin, Z.J. Tu, C.S. Gong, Y. Fu, S.H. Yan, H.C. Lei, Superconductivity and normal-state properties of Kagome metal RbV$_3$Sb$_5$ single crystals, Chin. Phys. Lett., 38 (2021) 037403.

[56] B.R. Ortiz, S.M.L. Teicher, Y. Hu, J.L. Zuo, P.M. Sarte, E.C. Schueller, A.M.M. Abeykoon, M.J. Krogstad, S. Rosenkranz, R. Osborn, R. Seshadri, L. Balents, J. He, S.D. Wilson, CsV$_3$Sb$_5$: A Z$_2$ topological Kagome metal with a superconducting ground state, Physical Review Letters, 125 (2020) 247002.

[57] Hong Li, Siyu Cheng, Brenden R. Ortiz, Hengxin Tan, Dominik Werhahn, Keyu Zeng, Dirk Jorhendt, Binghai Yan, Ziqiang Wang, Stephen D. Wilson, I. Zeljkovic, Electronic nematicity in the absence of charge density waves in a new titanium-based kagome metal, arXiv:2211.16477v1, DOI (2022).

[58] Haitao Yang, Zhen Zhao, Xin-Wei Yi, Jiali Liu, Jing-Yang You, Yuhang Zhang, Hui Guo, Xiao Lin, Chengmin Shen, Hui Chen, Xiaoli Dong, Gang Su, H.-J. Gao, Titanium-based kagome superconductor CsTi$_3$Bi$_5$ and topological states, arXiv:2209.03840v1, DOI.

[59] Haitao Yang, Yuhan Ye, Zhen Zhao, Jiali Liu, Xin-Wei Yi, Yuhang Zhang, Jinan Shi, Jing-Yang You, Zihao Huang, Bingjie Wang, Jing Wang, Hui Guo, Xiao Lin, Chengmin Shen, Wu Zhou, Hui Chen, Xiaoli Dong, Gang Su, Ziqiang Wang, H.-J. Gao, Superconductivity and orbital-selective nematic order in a new titaniumbased kagome metal CsTi$_3$Bi$_5$, arXiv:2211.12264v1, DOI (2022).

[60] L.C. Ying Zhou, Xuecong Ji, Chen Liu, Ke Liao, Zhongnan Guo, Jia'ou Wang, Hongming Weng, Gang Wang, Physical properties, electronic structure, and strain-tuned monolayer of the weak topological insulator RbTi$_3$Bi$_5$ with Kagome lattice, arXiv:2301.01633, DOI.

[61] F. Du, S. Luo, B.R. Ortiz, Y. Chen, W. Duan, D. Zhang, X. Lu, S.D. Wilson, Y. Song, H. Yuan, Pressure-induced double superconducting domes and charge instability in the kagome metal KV$_3$Sb$_5$, Phys. Rev. B, 103 (2021) L220504.

[62] K.Y. Chen, N.N. Wang, Q.W. Yin, Y.H. Gu, K. Jiang, Z.J. Tu, C.S. Gong, Y. Uwatoko, J.P. Sun, H.C. Lei, J.P. Hu, J.G. Cheng, Double superconducting dome and triple enhancement of T$_c$ in the Kagome superconductor CsV$_3$Sb$_5$ under high pressure, Physical Review Letters, 126 (2021) 247001.

[63] Y.M. Oey, B.R. Ortiz, F. Kaboudvand, J. Frassineti, E. Garcia, R. Cong, S. Sanna, V.F. Mitrović, R. Seshadri, S.D. Wilson, Fermi level tuning and double-dome superconductivity in the kagome metal CsV$_3$Sb$_{5-x}$Sn$_x$, Phys. Rev. Materials, 6 (2022) L041801.

[64] H. Yang, Z. Huang, Y. Zhang, Z. Zhao, J. Shi, H. Luo, L. Zhao, G. Qian, H. Tan, B. Hu, K. Zhu, Z. Lu, H. Zhang, J. Sun, J. Cheng, C. Shen, X. Lin, B. Yan, X. Zhou, Z. Wang, S.J. Pennycook, H. Chen, X. Dong, W. Zhou, H.-J. Gao, Titanium doped kagome superconductor CsV$_{3-x}$Ti$_x$Sb$_5$ and two distinct phases, Sci. Bull., 67 (2022) 2176-2185.

[65] Y.M. Oey, F. Kaboudvand, B.R. Ortiz, R. Seshadri, S.D. Wilson, Tuning charge density wave order and superconductivity in the kagome metals KV$_3$Sb$_{5-x}$Sn$_x$ and RbV$_3$Sb$_{5-x}$Sn$_x$, Phys. Rev. Materials, 6 (2022) 074802.

[66] X. Chen, X. Zhan, X. Wang, J. Deng, X.-B. Liu, X. Chen, J.-G. Guo, X. Chen, Highly Robust Reentrant Superconductivity in CsV$_3$Sb$_5$ under Pressure, Chin. Phys. Lett., 38 (2021) 057402.

[67] Y. Song, T. Ying, X. Chen, X. Han, X. Wu, A.P. Schnyder, Y. Huang, J. Guo, X. Chen, Competition of Superconductivity and Charge Density Wave in Selective Oxidized CsV$_3$Sb$_5$ Thin Flakes, Physical Review Letters, 127 (2021) 237001.

[68] G.P. Brenden R. Ortiz, Malia Gundayao, Hong Li, Farnaz Kaboudvand, Linus Kautzsch, Suchismita Sarker, Jacob P. C. Ruff, Tom Hogan, Steven J. Gomez Alvarado, Paul M. Sarte, Guang Wu, Tara Braden,

Ram Seshadri, Eric S. Toberer, Ilija Zeljkovic, Stephen D. Wilson, YbV$_3$Sb$_4$ and EuV$_3$Sb$_4$, vanadium-based kagome metals with Yb$^{2+}$ and Eu$^{2+}$ zig-zag chains, arXiv:2302.12354, DOI.

[69] A. Ovchinnikov, S. Bobev, Synthesis, Crystal and Electronic Structure of the Titanium Bismuthides Sr$_5$Ti$_{12}$Bi$_{19+x}$, Ba$_5$Ti$_{12}$Bi$_{19+x}$, and Sr$_{5-\delta}$Eu$_\delta$Ti$_{12}$Bi$_{19+x}$ (x ≈ 0.5–1.0; δ ≈ 2.4, 4.0), European Journal of Inorganic Chemistry, 2018 (2018) 1266-1274.

[70] P.C. Canfield, T. Kong, U.S. Kaluarachchi, N.H. Jo, Use of frit-disc crucibles for routine and exploratory solution growth of single crystalline samples, Philosophical Magazine, 96 (2016) 84-92.

[71] J. Rodríguez-Carvajal, FullProf, CEA/Saclay, France, DOI (2001).

[72] O.V. Dolomanov, L.J. Bourhis, R.J. Gildea, J.A.K. Howard, H. Puschmann, OLEX2: a complete structure solution, refinement and analysis program, J. Appl. Crystallogr., 42 (2009) 339-341.

[73] V. Petříček, M. Dušek, L. Palatinus, Crystallographic Computing System JANA2006: General features, Zeitschrift für Kristallographie - Crystalline Materials, 229 (2014) 345-352.

[74] C. Pfleiderer, S.R. Julian, G.G. Lonzarich, Non-Fermi-liquid nature of the normal state of itinerant-electron ferromagnets, Nature, 414 (2001) 427-430.

[75] D. van der Marel, J.L.M. van Mechelen, I.I. Mazin, Common Fermi-liquid origin of T$^2$ resistivity and superconductivity in n-type SrTiO$_3$, Phys. Rev. B, 84 (2011) 205111.

[76] W.G. Baber, N.F. Mott, The contribution to the electrical resistance of metals from collisions between electrons, Proceedings of the Royal Society of London. Series A - Mathematical and Physical Sciences, 158 (1937) 383-396.

[77] A.A. Abrikosov, I.M. Khalatnikov, THEORY OF THE FERMI FLUID (The Properties of Liquid He3 at Low Temperatures), Soviet Physics Uspekhi, 1 (1958) 68.

[78] N.F. Mott, Electrons in transition metals, Advances in Physics, 13 (1964) 325-422.

[79] H. Jiang, J.-K. Bao, H.-F. Zhai, Z.-T. Tang, Y.-L. Sun, Y. Liu, Z.-C. Wang, H. Bai, Z.-A. Xu, G.-H. Cao, Physical properties and electronic structure of Sr$_2$Cr$_3$As$_2$O$_2$ containing CrO$_2$ and Cr$_2$As$_2$ square-planar lattices, Phys. Rev. B, 92 (2015) 205107.

[80] A.H. Wilson, R.H. Fowler, The electrical conductivity of the transition metals, Proceedings of the Royal Society of London. Series A. Mathematical and Physical Sciences, 167 (1938) 580-593.

[81] X.-N. Luo, C. Dong, S.-K. Liu, Z.-P. Zhang, A.-L. Li, L.-H. Yang, X.-C. Li, Low-temperature physical properties and electronic structures of Ni$_3$Sb, Ni$_5$Sb$_2$, NiSb$_2$, and NiSb, Chin. Phys. B, 24 (2015) 067201.

[82] L. Chen, L.L. Zhao, X.L. Qiu, Q.H. Zhang, K. Liu, Q.S. Lin, G. Wang, Quasi-one-dimensional structure and possible helical antiferromagnetism of RbMn$_6$Bi$_5$, Inorganic Chemistry, 60 (2021) 12941-12949.

[83] Y. Zhou, L. Chen, G. Wang, Y.X. Wang, Z.C. Wang, C.C. Chai, Z.N. Guo, J.P. Hu, X.L. Chen, A new superconductor parent compound NaMn$_6$Bi$_5$ with quasi-one-dimensional structure and lower antiferromagnetic-like transition temperatures, Chin. Phys. Lett., 39 (2022) 047401.

[84] Y. Zhou, L. Chen, Y.X. Wang, J.F. Zhu, Z.N. Guo, C. Liu, Z.Y. Guo, C.W. Wang, H. Zhang, Y.L. Wang, K. Liao, Y.T. Song, J.O. Wang, D.L. Chen, J. Ma, J.P. Hu, G. Wang, ANi$_5$Bi$_{5.6+\delta}$ (A = K, Rb, and Cs): Quasi-One-Dimensional Metals Featuring [Ni$_5$Bi$_{5.6+\delta}$]$^-$ Double-Walled Column with Strong Diamagnetism, Inorganic Chemistry, 62 (2023) 3788-3798.

[85] N.W.a.M. Ashcroft, N.D., Solid State Physics, Saunders College, Philadelphia, 1976.

[86] Y. Li, E. Wang, X. Zhu, H.-H. Wen, Pressure-induced superconductivity in Bi single crystals, Phys. Rev. B, 95 (2017) 024510.

# Supporting Information for Publication
# Tunable magnetism and electron correlation in Titanium-based Kagome metals RETi$_3$Bi$_4$ (RE = Yb, Pr, and Nd) by rare-earth engineering

Long Chen[1,2,#], Ying Zhou[1,2,#], He Zhang[1,2,#], Xuecong Ji[1,2,#], Ke Liao[1,2], Yu Ji[1,2], Ying Li[1], Zhongnan Guo[3], Xi Shen[1], Richeng Yu[1,2,4], Xiaohui Yu[1,2,4,*], Hongming Weng[1,2,4,*], Gang Wang[1,2,4,*]

[1] Beijing National Laboratory for Condensed Matter Physics, Institute of Physics, Chinese Academy of Sciences, Beijing 100190, China

[2] University of Chinese Academy of Sciences, Beijing 100049, China

[3] Department of Chemistry, School of Chemistry and Biological Engineering, University of Science and Technology Beijing, Beijing 100083, China

[4] Songshan Lake Materials Laboratory, Dongguan, Guangdong 523808, China

[#] These authors contributed equally to this work.

*Corresponding author. Email: yuxh@iphy.ac.cn; hmweng@iphy.ac.cn; gangwang@iphy.ac.cn.

**Table S1**. Crystallographic data and structure refinement of RETi$_3$Bi$_4$ (RE = Yb, Pr and Nd).

| Empirical formula | | YbTi$_3$Bi$_4$** | PrTi$_3$Bi$_4$ | NdTi$_3$Bi$_4$ |
|---|---|---|---|---|
| f.u. weight (g/mol) | | 1152.7 | 1120.6 | 1120.6 |
| Space group / Z | | *Fmmm* (No.69) / 4 | | |
| Unit cell parameter | *a* (Å) | 24.95(4) | 24.9668(114) | 24.9523(87) |
| | *b* (Å) | 10.3(4) | 10.3248(49) | 10.3327(27) |
| | *c* (Å) | 5.9(2) | 5.9125(26) | 5.9009(18) |
| | *α, β, γ* (°) | 90 | 90 | 90 |
| | V (Å$^3$) | 1513.17(3) | 1524.108(2.0) | 1521.4 (1.9) |
| $d_{cal}$ (g/cm$^3$) | | -- | 4.883 | 4.883 |
| Refl. Collectd / unique | | -- | 2477 / 364 | 2013 / 346 |
| $R_{int}$ | | -- | 0.0723 | 0.0623 |
| Goodness-of-fit | | -- | 1.5212 | 1.435 |
| $R_1$ / $wR_2$ ($I > 2\sigma(I)$) | | -- | 0.0473 / 0.0887 | 0.0461 / 0.0884 |
| $R_1$ / $wR_2$ (all) | | -- | 0.0723 / 0.0989 | 0.0644 / 0.0963 |

** Due to the two dimensional feature of YbTi$_3$Bi$_4$, the collected SCXRD data cannot be used to do the refinement. Here we only used the indexed unit cell parameters. The actual crystal structure of YbTi$_3$Bi$_4$ is determined by scanning transmission electron microscope as shown in the manuscript.

**Table S2**. Atomic coordinates and equivalent isotropic displacement parameters for PrTi$_3$Bi$_4$.

| Atom | *Wyck.* | *Sym.* | x/a | y/b | z/c | Occ. | U(eq)(Å$^2$) |
|---|---|---|---|---|---|---|---|
| Pr | *8g* | *2mm* | 0.6954(1) | 0 | 0 | 1.0 | 0.0262(8) |
| Ti1 | *8g* | *m* | 0.5926(3) | 0 | -0.5000 | 1.0 | 0.0250(2) |
| Ti2 | *16l* | *m* | 0.5933(3) | 0.2500 | -0.2500 | 1.0 | 0.0290(2) |
| Bi1 | *8h* | *m* | 0.5000 | 0.1693(1) | -0.5000 | 1.0 | 0.0263(6) |
| Bi2 | *16o* | *m* | 0.6877(1) | 0.1613(1) | -0.5000 | 1.0 | 0.0263(4) |
| Bi3 | *8g* | *m* | 0.5682(1) | 0.5000 | 0.5000 | 1.0 | 0.0271(6) |

**Table S3**. Anisotropic displacement parameters for PrTi$_3$Bi$_4$. The anisotropic displacement factor exponent takes the form: $-2\pi^2[h^2a^2U_{11} + ... + 2hkabU_{12}]$.

| Atom | $U_{11}$ (Å$^2$) | $U_{22}$ (Å$^2$) | $U_{33}$ (Å$^2$) | $U_{23}$ (Å$^2$) | $U_{13}$ (Å$^2$) | $U_{12}$ (Å$^2$) |
|---|---|---|---|---|---|---|
| Pr | 0.0480(18) | 0.0076(7) | 0.0230(13) | 0 | 0 | 0 |
| Ti1 | 0.0400(60) | 0.0080(20) | 0.0280(40) | 0 | 0 | 0 |
| Ti2 | 0.0630(50) | 0.0083(17) | 0.0160(30) | 0 | 0 | -0.0021(14) |
| Bi1 | 0.0462(13) | 0.0080(50) | 0.0246(10) | 0 | 0 | 0 |
| Bi2 | 0.0493(10) | 0.0071(4) | 0.0226(7) | -0.0002(3) | 0 | 0 |
| Bi3 | 0.0568(14) | 0.0053(5) | 0.0192(8) | 0 | 0 | 0 |

**Table S4**. Atomic coordinates and equivalent isotropic displacement parameters for NdTi$_3$Bi$_4$.

| Atom | Wyck. | Sym. | x/a | y/b | z/c | Occ. | U(eq)(Å$^2$) |
|------|-------|------|-----|-----|-----|------|--------------|
| Nd   | 8g    | 2mm  | 0.30408(8)  | 0         | 0.5000 | 1.0 | 0.0142(6) |
| Ti1  | 8g    | m    | 0.4071(3)   | 0.5000    | 0.5000 | 1.0 | 0.0140(20) |
| Ti2  | 16l   | m    | 0.4053(2)   | 0.7500    | 0.7500 | 1.0 | 0.0192(17) |
| Bi1  | 8h    | m    | 0.5000      | 0.6694(1) | 0.5000 | 1.0 | 0.0143(5) |
| Bi2  | 16o   | m    | 0.3118(1)   | 0.6606(1) | 0.5000 | 1.0 | 0.0139(4) |
| Bi3  | 8g    | m    | 0.43123(5)  | 0         | 0.5000 | 1.0 | 0.0144(5) |

**Table S5**. Anisotropic displacement parameters for NdTi$_3$Bi$_4$. The anisotropic displacement factor exponent takes the form: $-2\pi^2[h^2a^2U_{11} + ... + 2hkabU_{12}]$.

| Atom | $U_{11}$ (Å$^2$) | $U_{22}$ (Å$^2$) | $U_{33}$ (Å$^2$) | $U_{23}$ (Å$^2$) | $U_{13}$ (Å$^2$) | $U_{12}$ (Å$^2$) |
|------|-----------|-----------|-----------|-----------|-----------|-----------|
| Nd   | 0.0202(12) | 0.0104(8)  | 0.0119(12) | 0         | 0 | 0 |
| Ti1  | 0.029(5)   | 0.011(3)   | 0.003(4)   | 0         | 0 | 0 |
| Ti2  | 0.047(4)   | 0.0047(17) | 0.006(3)   | 0         | 0 | 0.0013(17) |
| Bi1  | 0.0214(9)  | 0.0076(5)  | 0.0138(9)  | 0         | 0 | 0 |
| Bi2  | 0.0215(8)  | 0.0006(2)  | 0.0072(5)  | 0.0131(7) | 0 | 0 |
| Bi3  | 0.0253(9)  | 0.0051(5)  | 0.0127(9)  | 0         | 0 | 0 |

**Table S6.** Quantitative analysis of elemental composition for YbTi$_3$Bi$_4$ single crystals. For each sample, the molar ratio of composition elements is measured at three different locations. The stoichiometry of YbTi$_3$Bi$_4$ is determined to be Yb : Ti : Bi = 0.85(3) : 3.00(5) : 4.04(3) by the statistical average and standard deviation (SD) of the overall energy-dispersive spectroscopy (EDS) data.

| Atom    | Yb (Mol%) | Ti (Mol%) | Bi (Mol%) |
|---------|-----------|-----------|-----------|
| SA#-1   | 10.62     | 38.02     | 51.36     |
| SA#-2   | 11.37     | 37.37     | 51.26     |
| SA#-3   | 10.74     | 37.74     | 51.51     |
| SB#-1   | 11.07     | 37.03     | 51.90     |
| SB#-2   | 10.30     | 39.07     | 50.63     |
| SB#-3   | 10.74     | 37.90     | 51.36     |
| SC#-1   | 11.39     | 37.78     | 50.83     |
| SC#-2   | 10.26     | 38.65     | 51.08     |
| SC#-3   | 10.79     | 38.29     | 50.93     |
| Average | 10.81     | 37.98     | 51.21     |
| SD      | 0.41      | 0.62      | 0.39      |

**Table S7.** Quantitative analysis of elemental composition for PrTi$_3$Bi$_4$ single crystals. The final stoichiometry of PrTi$_3$Bi$_4$ is determined to be Pr : Ti : Bi = 1.00(3) : 3.06(7) : 3.99(8).

| Atom | Pr (Mol%) | Ti (Mol%) | Bi (Mol%) |
|---|---|---|---|
| SA#-1 | 12.82 | 37.61 | 49.57 |
| SA#-2 | 11.79 | 38.65 | 49.56 |
| SA#-3 | 12.33 | 36.42 | 51.25 |
| SB#-1 | 12.28 | 36.98 | 50.74 |
| SB#-2 | 12.87 | 38.66 | 48.47 |
| SB#-3 | 12.72 | 39.03 | 48.25 |
| SC#-1 | 11.99 | 38.99 | 49.02 |
| SC#-2 | 12.92 | 37.49 | 49.59 |
| SC#-3 | 12.12 | 38.28 | 49.60 |
| Average | 12.43 | 38.01 | 49.56 |
| SD | 0.42 | 0.93 | 0.97 |

**Table S8.** Quantitative analysis of elemental composition for NdTi$_3$Bi$_4$ single crystals. The final stoichiometry of NdTi$_3$Bi$_4$ is determined to be Nd : Ti : Bi = 1.00(4) : 3.02(5) : 3.92(5).

| Atom | Nd (Mol%) | Ti (Mol%) | Bi (Mol%) |
|---|---|---|---|
| SA#-1 | 13.14 | 37.81 | 49.05 |
| SA#-2 | 11.69 | 37.96 | 50.35 |
| SA#-3 | 12.27 | 38.72 | 49.01 |
| SB#-1 | 12.53 | 37.20 | 50.27 |
| SB#-2 | 11.94 | 38.58 | 49.48 |
| SB#-3 | 12.55 | 37.93 | 49.52 |
| SC#-1 | 12.93 | 38.57 | 48.50 |
| SC#-2 | 13.18 | 38.41 | 48.41 |
| SC#-3 | 13.10 | 37.24 | 49.66 |
| Average | 12.59 | 38.05 | 49.36 |
| SD | 0.54 | 0.57 | 0.69 |

**Table S9**. The fitted parameters of magnetic susceptibility for RETi$_3$Bi$_4$ using Curie-Weiss law.

| RE | Yb | | Pr | | Nd | |
|---|---|---|---|---|---|---|
| Field configuration | $H // bc$ | $H \perp bc$ | $H // bc$ | $H \perp bc$ | $H // bc$ | $H \perp bc$ |
| $\chi_0$ (10$^{-4}$ emu mol$^{-1}$ Oe$^{-1}$) | 0.36(1) | 2.29(6) | -6.50(4) | -32.2(1) | 0.21(4) | 2.17(2) |
| $C$ (10$^{-3}$ emu mol$^{-1}$ Oe$^{-1}$ K$^{-1}$) | 9.59(5) | 3.11(2) | 1.68(5) | 2.78(8) | 1.58(3) | 1.70(1) |
| $\theta$ (K) | 0.55(1) | 0.55(1) | 2.50(1) | -100(1) | -3.62(3) | -36.7(3) |
| $\mu_{eff}$ ($\mu_B$/f.u.) | 0.28(1) | 0.16(1) | 3.66(5) | 4.72(7) | 3.58(6) | 3.69(1) |
| $\mu_{sat}$ at 7 T ($\mu_B$/ f.u.) | 0.016 | 0.006 | 1.620 | 0.855 | 1.426 | 0.587 |

**Table S10**. The fitted parameters of resistivity using power law and that of specific heat capacity using Debye model for RETi$_3$Bi$_4$.

| RE | Yb | Pr | Nd | Nd |
|---|---|---|---|---|
| $T_{anomaly}$ or $T_c$ | - | ~8.2 K | 8.5 K | 8.5 K |
| $T_{min}$ - $T_{max}$ | 2 K - 45 K | 10 K - 45 K | 10 K - 45 K | 2 K - 9 K |
| $\rho_0$ (10$^{-5}$ Ω cm) | 2.62(1) | 3.68(3) | 3.79(1) | 3.45(1) |
| $A$ (10$^{-9}$ Ω cm K$^{-\alpha}$) | 2.36(29) | 8.07(1.02) | 2.56(28) | 0.00329(6) |
| $\alpha$ | 2.00(3) | 1.64(3) | 1.51(2) | 5.35(8) |
| $T_{min}$ - $T_{max}$ | 2 K - 7 K | 12 K - 18 K | 12 K - 18 K | |
| $\gamma$ (mJ mol$^{-1}$ K$^{-2}$) | 20.7(7) | 762.6(8.9) | 809.6(7.5) | |

**Table S11**. The fitted parameters of resistance for NdTi$_3$Bi$_4$ under high pressure using power law (10 K - 45 K).

| Pressure (GPa) | $R_0$ (Ω) | $A$ (10$^{-5}$ Ω K$^{-\alpha}$) | $\alpha$ |
|---|---|---|---|
| 1.5 | 0.06853(4) | 3.65(17) | 1.539(12) |
| 2.6 | 0.06900(2) | 3.75(9) | 1.530(6) |
| 3.8 | 0.06924(2) | 3.09(7) | 1.546(6) |
| 5.4 | 0.07044(2) | 3.12(10) | 1.544(8) |
| 7.7 | 0.07141(3) | 2.97(13) | 1.554(11) |
| 10.2 | 0.07180(3) | 2.89(13) | 1.556(12) |
| 12.4 | 0.07111(4) | 2.48(15) | 1.594(16) |
| 15.4 | 0.07237(4) | 2.03(14) | 1.640(18) |
| 18.9 | 0.07585(4) | 1.07(10) | 1.802(26) |
| 23.3 | 0.07975(4) | 1.12(11) | 1.794(26) |
| 28.1 | 0.08467(4) | 1.02(10) | 1.818(25) |
| 32.4 | 0.08675(5) | 9.21(10) | 1.854(28) |
| 37.0 | 0.08938(5) | 1.19(13) | 1.810(28) |
| 41.9 | 0.09137(6) | 1.42(15) | 1.786(27) |

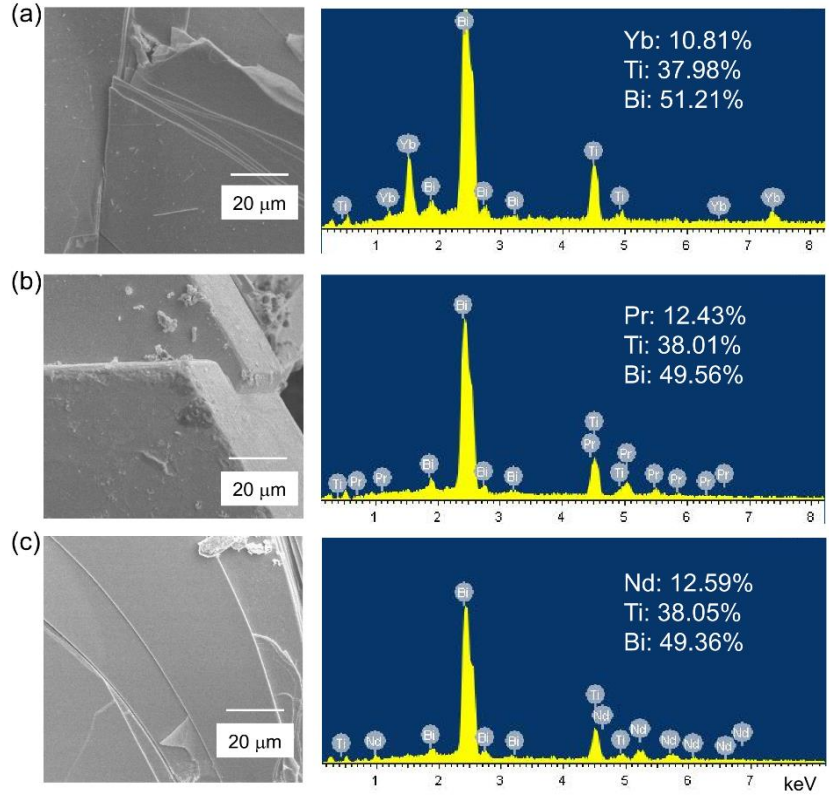

**Figure S1.** SEM images and typical EDSs collected on the flat clean surface of (a) YbTi$_3$Bi$_4$, (b) PrTi$_3$Bi$_4$, and (c) NdTi$_3$Bi$_4$ single crystals, showing clean surfaces with hexagonal shape and two-dimensional feature.

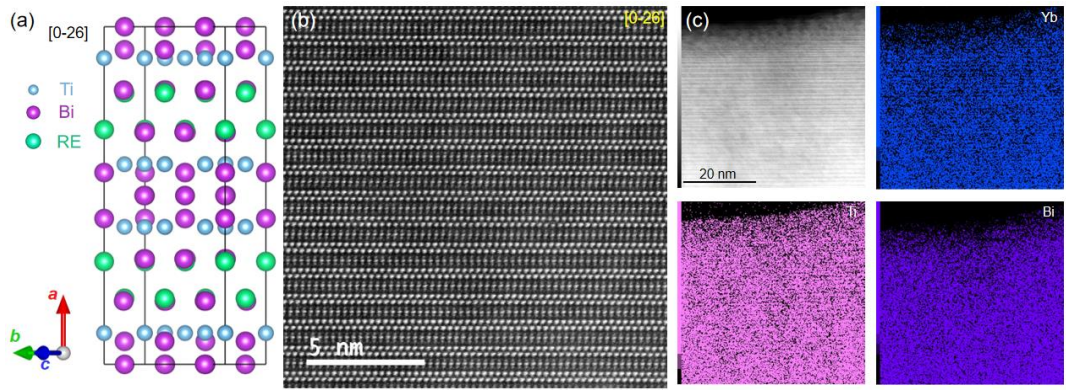

**Figure S2.** The crystal structure and (b) high-angle annular dark-field image of YbTi$_3$Bi$_4$ along the [0-26] zone-axis. (c) EDS mapping of YbTi$_3$Bi$_4$ collected on a 20 nm scale bar, showing the homogenous distribution of Yb, Ti and Bi elements

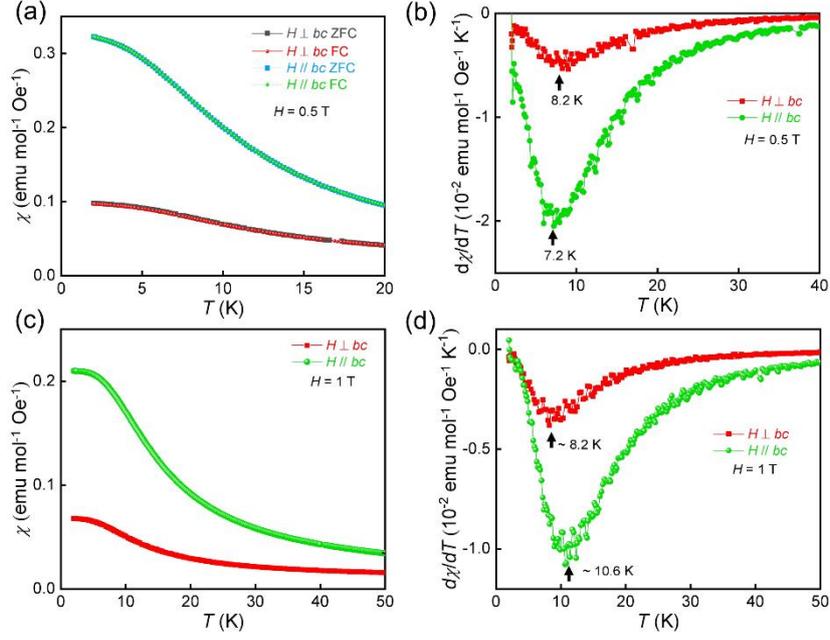

**Figure S3.** (a) Magnetic susceptibility and (b) first derivative of magnetic susceptibility ($d\chi/dT$) for PrTi$_3$Bi$_4$ at low temperature under 0.5 T. (c) Magnetic susceptibility and (d) first derivative of magnetic susceptibility ($d\chi/dT$) for PrTi$_3$Bi$_4$ at low temperature under 1 T.

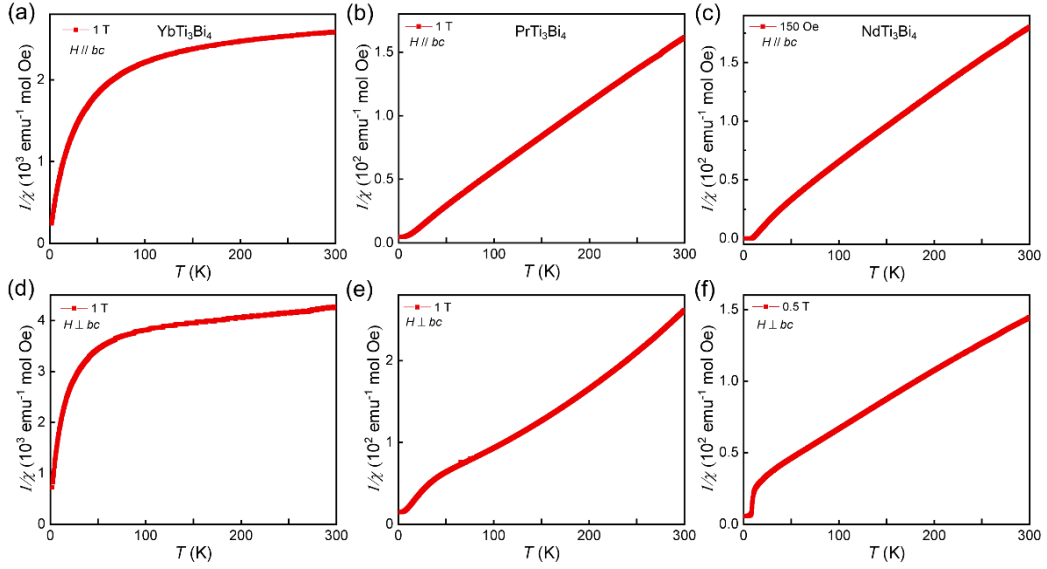

**Figure S4.** $1/\chi$ vs $T$ plot with magnetic field parallel to the $bc$ plane of (a) YbTi$_3$Bi$_4$, (b) PrTi$_3$Bi$_4$, and (c) NdTi$_3$Bi$_4$ single crystals. $1/\chi$ vs $T$ plot with magnetic field perpendicular to the $bc$ plane of (a) YbTi$_3$Bi$_4$, (b) PrTi$_3$Bi$_4$, and (c) NdTi$_3$Bi$_4$ single crystals.

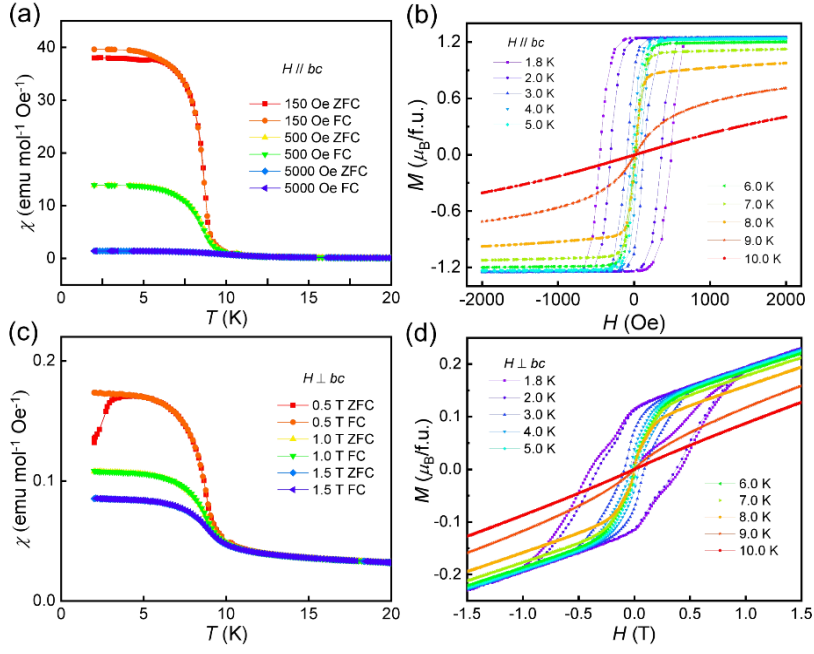

**Figure S5.** Magnetic susceptibility of NdTi$_3$Bi$_4$ with different magnetic field (a) parallel and (c) perpendicular to the *bc* plane. Field-dependent magnetization of NdTi$_3$Bi$_4$ with magnetic field (b) parallel and (d) perpendicular to the *bc* plane.

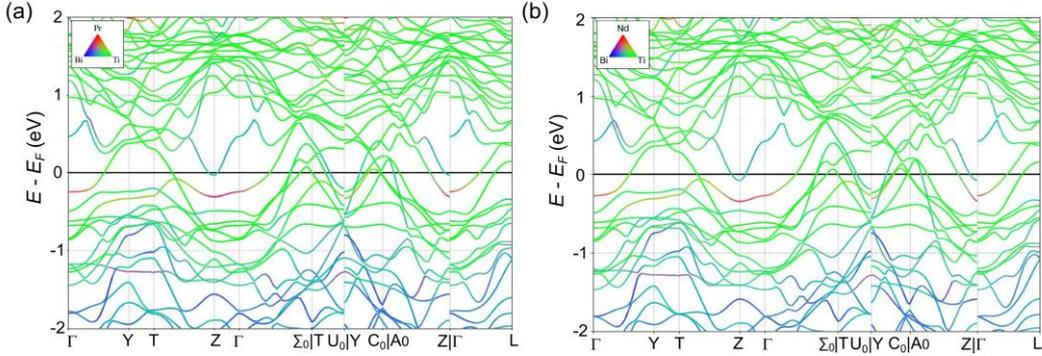

**Figure S6.** The metallic electronic structure of (a) PrTi$_3$Bi$_4$ and (b) NdTi$_3$Bi$_4$. The different color of bands represent the contribution coming from Pr or Nd (red), Ti (green) and Bi (blue).

First-principles calculations were carried out with the projector augmented wave method as implemented in the Vienna *ab initio* simulation Package [1-3]. The generalized gradient approximation [4] of the Perdew-Burke-Ernzerhof [2] type was adopted for the exchange-correlation function. The cutoff energy of the plane-wave basis was 500 eV and the energy convergence standard was set to $10^{-8}$ eV. The $3 \times 11 \times 4$ Monkhorst-Pack K-point mesh was employed for the Brillouin zone (BZ) sampling of the $1 \times 1 \times 1$ unit cell. Since Bi is a quite heavy element, spin-orbit coupling (SOC) effect is taken into account by treating core-electrons in fully relativistic method and valance electrons in second-variation method.

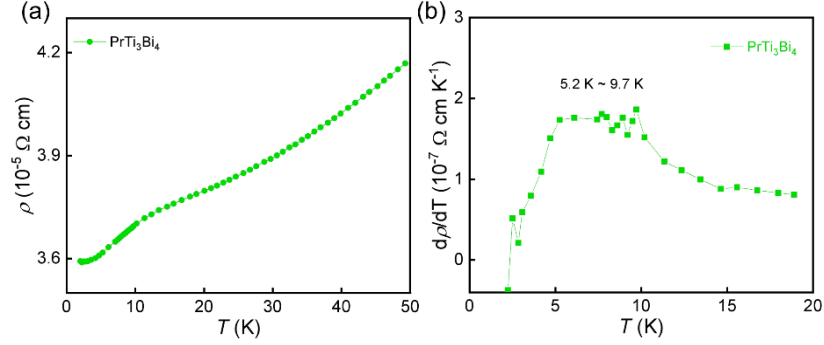

**Figure S7.** Zoomed temperature-dependent resistivity of PrTi$_3$Bi$_4$ at low temperature. The arrow denotes the abnormal enhancement around 10 K.

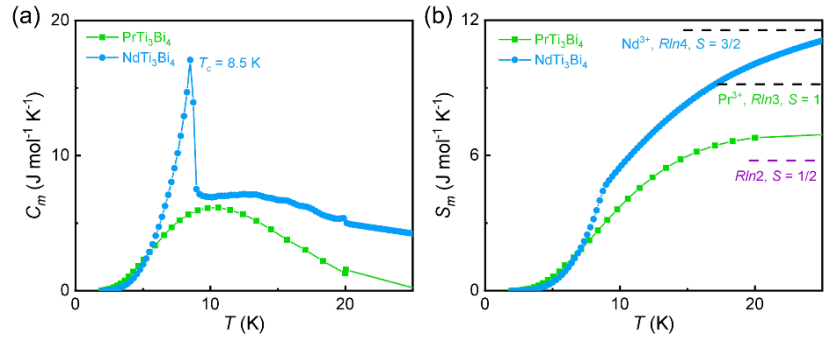

**Figure S8.** (a) $C_{mag}/T$ vs $T$ and (b) the calculated magnetic entropy $S_{mag}$ of PrTi$_3$Bi$_4$ and NdTi$_3$Bi$_4$ single crystals. The dashed lines show the theoretical value $Rln(2S + 1)$ with $S = 1/2$, 1, and 3/2.

In order to obtain the specific heat capacity contributed by magnon ($C_{mag}$) associated with the broaden anomaly of PrTi$_3$Bi$_4$ and FM phase transition of NdTi$_3$Bi$_4$, the specific heat capacity of nonmagnetic YbTi$_3$Bi$_4$ is taken as the sum of phonon contribution ($C_{ph}$) and electron contribution ($C_e$), $C_{Yb} = C_{ph} + C_e$. By subtracting the specific heat capacity of YbTi$_3$Bi$_4$ from the total heat capacity ($C_p$) of PrTi$_3$Bi$_4$ and NdTi$_3$Bi$_4$, the specific heat capacity contributed by magnon is calculated using $C_{mag} = C_p - C_{Yb} = C_p - C_{ph} - C_e$. The magnetic entropy is further calculated following $S_{mag}(T) = \int_0^T C_{mag}/T dT$. For PrTi$_3$Bi$_4$, the specific heat capacity contributed by magnons show a broaden peak around 10 K. The magnetic entropy tends to saturated at 25 K and exceeds $Rln2 = 5.78$ J mol$^{-1}$ K$^{-1}$ with $S = 1/2$, possibly indicating a Pr$^{3+}$ state with $S = 1$. For NdTi$_3$Bi$_4$, a sharp peak at 8.5 K can be observed in its specific heat capacity contributed by magnon and the saturated magnetic entropy approach the theoretical value $Rln(2S+1) = 11.56$ J mol$^{-1}$ K$^{-1}$ for $S = 3/2$, suggesting a Nd$^{3+}$ state.

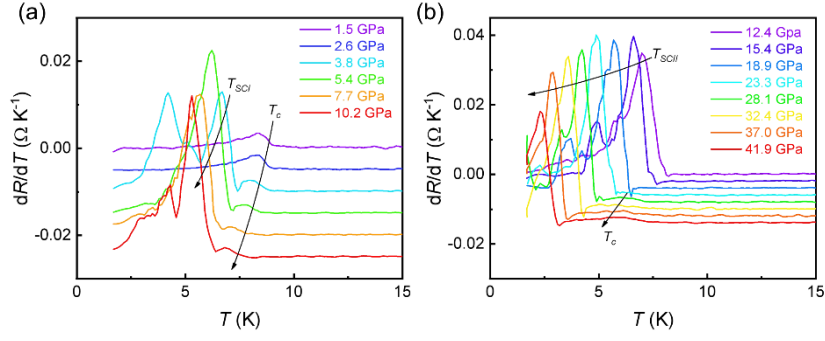

**Figure S9.** The first derivative curves of resistance (d$R$/d$T$) for NdTi$_3$Bi$_4$ under high pressure from (a) 1.5 GPa to 10.2 GPa, (b) 12.4 GPa to 41.9 GPa. The arrows show the suppression trend of ferromagnetic ordering temperature ($T_c$) and superconducting transition temperature ($T_{SCI}$ or $T_{SCII}$).

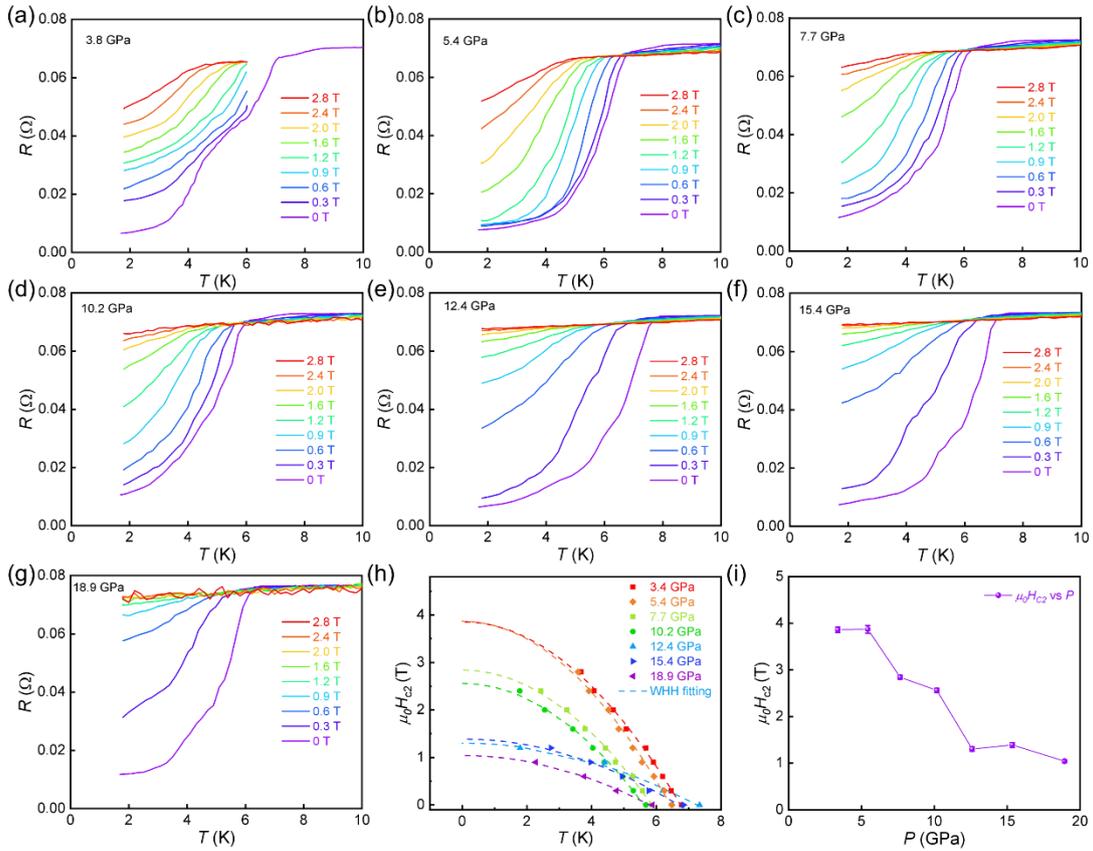

**Figure S10.** Resistance measurements under different magnetic field for the pressurized NdTi$_3$Bi$_4$ under (a) 3.8 GPa, (b) 5.4 GPa, (c) 7.7 GPa, (d) 10.2 GPa, (e) 12.4 GPa, (f) 15.4 GPa, and (g) 18.9 GPa. (h) Plots of superconducting transition temperature ($T_{SC}$) vs upper critical field ($\mu_0H_{c2}$) for NdTi$_3$Bi$_4$ under different pressure. The dashed lines represent the WHH fits to the data of $\mu_0H_{c2}$. (i) Pressure dependence of upper critical field for NdTi$_3$Bi$_4$.

Owing to the non-zero resistance, the pressure-induced superconducting behavior may not be the intrinsic properties of NdTi$_3$Bi$_4$. In order to determine resolve this question, the resistance under different magnetic fields (0 T - 2.8 T) have been measured. As shown in Figure S9a-g, the drop in resistance is suppressed with applied magnetic field. Without magnetic field under 3.8 GPa, the superconducting transition shows a step-like feature with two superconducting transition, $T_{SC1}$ = 4.2

K and $T_{SC2}$ = 6.3 K, which is quite close to the reported pressure-induced superconducting Bi-II ($T_{SC}$ = 3.92 K) and Bi-III ($T_{SC}$ = 7 K) phases under 2.8 GPa [5], respectively. A rather small magnetic field (< 0.3 T) would fully suppress the lower superconducting transition but the higher superconducting transition would survive beyond 2.8 T. According to the Werthamer–Helfand–Hohenberg (WHH) theory [6], the upper critical field for the higher superconducting transition under 3.8 GPa is calculated to be 3.8 T, quite close to the reported upper critical field (3.71 T) for Bi-III phase [5]. For pressures from 5.4 GPa to 10.2 GPa, there only exists one superconducting transition, where both the superconducting transition temperature $T_{SC}$ and upper critical field slightly decrease with increasing pressure (Figure S9i), similar to the reported behavior of Bi-III phase [5]. When pressure increases to 12.4 GPa, another superconducting transition with higher $T_{SC}$ and lower upper critical field appears, which should be attributed to the reported Bi-V phase under high pressure [5].

[1] G. Kresse, J. Hafner, Abinitio molecular-dynamics for liquid-metals, Phys. Rev. B, 47 (1993) 558-561.
[2] G. Kresse, J. Furthmuller, Efficiency of ab-initio total energy calculations for metals and semiconductors using a plane-wave basis set, Computational Materials Science, 6 (1996) 15-50.
[3] G. Kresse, J. Furthmuller, Efficient iterative schemes for ab initio total-energy calculations using a plane-wave basis set, Phys. Rev. B, 54 (1996) 11169-11186.
[4] J.P. Perdew, K. Burke, M. Ernzerhof, Generalized gradient approximation made simple (vol 77, pg 3865, 1996), Physical Review Letters, 78 (1997) 1396-1396.
[5] Y. Li, E. Wang, X. Zhu, H.-H. Wen, Pressure-induced superconductivity in Bi single crystals, Phys. Rev. B, 95 (2017) 024510.
[6] N.R. Werthamer, E. Helfand, P.C. Hohenberg, Temperature and purity dependence of the superconducting critical field, $H_{c2}$. III. Electron spin and spin-orbit effects, Phys. Rev., 147 (1966) 295-302.